\documentclass[twocolumn,trackchanges]{aastex62}
\usepackage{amsmath,amstext}
\usepackage[T1]{fontenc}
\usepackage{natbib}

\received{date}
\revised{date}
\accepted{date}
\submitjournal{ApJ}

\shorttitle{New millimeter CO observations of the gas-rich debris disks 49\,Cet and HD\,32297}
\shortauthors{Mo\'or et al.}

\begin{document}

\title{New millimeter CO observations of the gas-rich debris disks 49\,Cet and HD\,32297}

\correspondingauthor{Attila Mo\'or}
\email{moor@konkoly.hu}

\author{Attila Mo\'or}
\affil{Konkoly Observatory, Research Centre for Astronomy and
Earth Sciences, Hungarian Academy of Sciences, Konkoly-Thege
Mikl\'os \'ut 15-17, 1121 Budapest, Hungary}

\author{Quentin Kral} 
\affiliation{LESIA, Observatoire de Paris, Universit\'e PSL, CNRS, Sorbonne Universit\'e, Univ. 
Paris Diderot, Sorbonne Paris Cit\'e, 5 place Jules Janssen, 92195 Meudon, France}

\author{P\'eter \'Abrah\'am} 
\affiliation{Konkoly Observatory,
Research Centre for Astronomy and Earth Sciences, Hungarian Academy
of Sciences, Konkoly-Thege Mikl\'os \'ut 15-17, 1121 Budapest,
Hungary}

\author{\'Agnes K\'osp\'al} \affiliation{Konkoly Observatory, Research
Centre for Astronomy and Earth Sciences, Hungarian Academy of
Sciences, Konkoly-Thege Mikl\'os \'ut 15-17, 1121 Budapest, Hungary}
\affiliation{Max Planck Institute for
Astronomy, K\"onigstuhl 17, D-69117 Heidelberg, Germany}

\author{Anne Dutrey} \affiliation{Laboratoire d'Astrophysique de Bordeaux, 
Univ. Bordeaux, CNRS, B18N, All\'ee Geoffroy Saint-Hilaire, 33615 Pessac, France}

\author{Emmanuel Di Folco} \affiliation{Laboratoire d'Astrophysique de Bordeaux, 
Univ. Bordeaux, CNRS, B18N, All\'ee Geoffroy Saint-Hilaire, 33615 Pessac, France}
  
\author{A. Meredith Hughes} \affiliation{Department of Astronomy, Van
Vleck Observatory, Wesleyan University, 96 Foss Hill Drive,
Middletown, CT 06459, USA}

\author{Attila Juh\'asz} \affiliation{Institute of Astronomy,
University of Cambridge, Madingley Road, Cambridge CB3 0HA, UK}

\author{Ilaria Pascucci} \affiliation{Lunar and Planetary Laboratory, The University of 
Arizona, Tucson, AZ 85721, USA}

\author{Nicole Pawellek} \affiliation{Max Planck Institute for
Astronomy, K\"onigstuhl 17, D-69117 Heidelberg, Germany}
\affiliation{Konkoly Observatory, Research
Centre for Astronomy and Earth Sciences, Hungarian Academy of
Sciences, Konkoly-Thege Mikl\'os \'ut 15-17, 1121 Budapest, Hungary}

\begin{abstract} 
Previous observations revealed the existence of CO gas at nearly protoplanetary level in several 
dust-rich debris disks around young A-type stars. Here we used the ALMA 7m-array to measure 
$^{13}$CO and C$^{18}$O emission toward two debris disks, 49\,Cet and HD\,32297, and 
detected similarly high CO content ($>$0.01\,M$_\oplus$). These high CO masses imply a highly efficient shielding 
of CO molecules against stellar and interstellar ultraviolet photons. Adapting a recent secondary gas 
disk model that considers both shielding by carbon atoms and self-shielding of CO, we can explain 
the observed CO level in both systems. Based on the derived gas densities we suggest that, 
in the HD\,32297 disk, dust and gas are coupled and the dynamics of small grains 
is affected by the gaseous component. For 49\,Cet, the question of coupling remains undecided. 
We found that the main stellar and disk properties
of 49\,Cet and HD\,32297 are very similar 
to those of previously identified debris disks with high CO content. 
These objects constitute together the first known representatives of shielded debris disks.

\end{abstract} 



\section{Introduction} \label{sec:intro} 
Circumstellar debris disks are mainly comprised of solids from micron to 
planetesimal size scales. Due to interactions with the stellar radiation field 
in such optically thin disks dust grains are rapidly removed. Thus the smallest 
dust particles -- that make these disks detectable for us at infrared (IR) wavelengths
via their thermal emission -- could not be leftovers from 
the primordial phase but should be second-generation grains stemming from 
collisional grinding of previously formed larger bodies \citep{wyatt2008}.
These parent planetesimals are confined into a belt or belts \citep{hughes2018}.
If icy planetesimals and grains are present, then their collisions and sublimation
may result in gas liberation as well \citep{zuckerman2012,kral2017}. Moreover, 
photodesorption of icy mantles can also contribute to gas release \citep{grigorieva2007}. 
Similarly to debris dust particles, gas produced in these ways is of second generation, rather than 
remnant gas from the protoplanetary phase.
  
Measuring the gas content of debris disks is crucial from several different points of 
view. Observation of different gas compounds offers a unique opportunity to constrain the 
volatile composition of the parent icy planetesimals \citep{kral2016,matra2017b,matra2018a}. 
Formation of rings, spirals, arcs in the distribution of small debris grains can either be 
caused  by the gravitational perturbation of an unseen planet or interactions with a sufficient 
amount of gas in the system \citep{takeuchi2001,richert2018}. To assess which mechanism could 
be active, one needs to measure the amount of gas. However, while the dust material of debris 
disks is rather well studied, their possible 
gas content remains less explored so far, especially due to observational difficulties. 
Nevertheless, intensive investigations 
over the past decades revealed gas emission in nineteen debris disks. Though in a few 
systems the gas component was detected through fine-structure lines of neutral oxygen or 
ionized carbon \citep{rm2012,rm2014}, so far  observation of low level rotational transitions
of $^{12}$CO molecules turned out to be the most effective way to probe gas in debris disks. 
Thanks to observations with single-dish radio telescopes and especially with the Atacama Large 
Millimeter/submillimeter Array (ALMA), as of today we know {seventeen} CO-bearing debris disks\footnote{The 
CO-bearing disk around the young pre-main-sequence Herbig B9.5 Ve star, HD\,141569, exhibits 
IR excess at only slightly higher level than that of the most dust-rich systems among the known 
 gaseous debris disks, suggesting a relationship to them \citep[e.g.][]{flaherty2016,pericaud2016}. However, 
 in several aspects this system differs from genuine debris disks \citep{wyatt2015,white2018}. 
Because of its not fully clarified evolutionary state we will handle this object separately from the 
CO-bearing debris disk sample.}.
Thirteen disks surround A-type stars: 49\,Cet \citep{zuckerman1995},  
HD\,21997 \citep{moor2011}, HD\,32297 \citep{greaves2016}, $\beta$\,Pic \citep{dent2014}, 
HD\,95086 \citep{booth2018}, HD\,121191, HD\,121617, HD\,131488 \citep{moor2017}, HD\,131835 \citep{moor2015b}, 
HD\,110058, HD\,138813, HD\,156623 \citep{lieman-sifry2016}, and Fomalhaut \citep{matra2017b}. 
Three disks are present in F-type systems, HD\,146897 \citep{lieman-sifry2016}, HD\,181327 \citep{marino2016}, 
$\eta$\,Crv \citep{marino2017} {and recently \citet{matra2019} revealed CO for the first time in a debris 
disk hosted by an M-type star, TWA\,7.}

The $^{12}$CO line J=2--1 luminosities of the detected gaseous debris disks show a 
large spread of almost four orders of magnitude, the brightest eight of them (49\,Cet, 
HD\,21997, HD\,32297, HD\,121617, HD\,131488, HD\,131835, HD\,138813, and HD\,156623) 
exhibit line luminosities higher than {$\sim$10$^{18}$\,W, i.e. 6x higher than that of $\beta$\,Pic 
\citep{matra2018a}.}
Besides their 
strong $^{12}$CO emission these eight disks share additional common properties. 
{They all belong to the most dust-rich known debris systems with 
their dust fractional luminosities of $f_d>$5$\times$10$^{-4}$ \citep[although 
there exist similarly dust-rich debris disks with no detectable CO, 
the most prominent example is HR\,4796,][]{kennedy2018}.}   
They surround young (10--50\,Myr) A-type stars, and the bulk 
of their detected gas material orbit at radial distances larger than $\sim$20\,au. 
Observations of the less abundant $^{13}$CO and C$^{18}$O isotopologues in the disks around 
HD\,21997, HD\,121617, HD\,131488 and HD\,131835 brought a surprising result: 
the measured line ratios implied highly optically thick $^{12}$CO emission 
in all cases. The total CO gas masses of these systems, based on the analysis 
of the optically thin C$^{18}$O line, are higher than 0.01\,$M_\oplus$ 
\citep{kospal2013,moor2017}.
The obtained CO masses are three to four orders of magnitude higher than those of 
the similarly young and dust-rich disks of $\beta$\,Pic and HD\,181327 
\citep{matra2017a,marino2016}, and resemble more the CO gas quantity 
of protoplanetary disks around Herbig Ae stars \citep{moor2017,kral2018}. 
  
Since the low dust content of debris disks does not provide efficient shielding 
against stellar and interstellar UV photons, gas molecules released from icy bodies 
are expected to be rapidly photodissociated. Unshielded second generation CO molecules 
are destroyed on a timescale of $\sim$100\,yrs by high energy photons of the interstellar 
radiation field \citep{visser2009}. This short lifetime is consistent with the low CO 
content of $\beta$\,Pic \citep{matra2017a} or HD\,181327 \citep{marino2016} but is 
in contradiction with the presence of the above-mentioned substantially more CO rich 
disks whose long term maintenance would require unrealistically high gas production rate. 
The existence of the latter disks can be understood if their CO molecules are shielded 
more effectively. The young age of these systems raises the 
possibility that the observed gas is not second generation but predominantly leftover 
from the primordial phase where remnant hydrogen molecules provide strong shielding to 
CO gas \citep{kospal2013}. Since the dust is thought to be second generation such disks 
would exhibit a hybrid nature \citep{kospal2013,pericaud2017}. 
As an alternative solution \citet{kral2018} recently demonstrated that shielding by neutral carbon 
gas (C$^0$) produced mainly from photodissociation of CO and CO$_2$ in tandem with self-shielding by 
CO molecules can increase the lifetime of CO gas released from planetesimals significantly.
They found that the gas quantity 
measured in all four above-mentioned CO-rich debris disks can be explained by applying their 
shielded secondary gas disk model without the presence of any primordial material.
  
{As part of our long term project to} study the properties and nature of {all known} young dust-rich 
debris disks exhibiting high $^{12}$CO line 
luminosity, in this work we investigated the CO content of two other representants, 49\,Cet 
and HD\,32297. In Section~\ref{sec:targets} we briefly summarize the currently known properties 
of our targets with special emphasis on previous studies regarding their gas material. 
Our new continuum and CO isotopologue observations carried out by the ALMA 7-m Array are 
described in Section~\ref{sec:obs}. Results from these observations including estimates for 
their dust and CO contents are presented in Section~\ref{sec:res}.
Implications of these results are discussed in Section~\ref{sec:discussion}.
Finally the outcomes of this investigation are summarized in Section~\ref{sec:summary}.

\section{Targets} \label{sec:targets} 
49\,Cet is an A1V-type star located at a distance of 57.0$\pm$0.3\,pc \citep[Gaia~DR2,][]{brown2018,lindegren2018,cb2018}. 
Based on its membership in the Argus moving group the estimated age 
of the star is 40--50\,Myr \citep{zuckerman2018}. The debris disk of 49\,Cet was discovered by \citet{sadakane1986} 
through its strong far-infrared excess. The spectral energy distribution (SED) of the excess 
is best described by a two-component model \citep{roberge2013} with characteristic dust 
temperatures of 165  and 59\,K and with a total fractional luminosity $f_d$=8.5$\times$10$^{-4}$ 
\citep{holland2017}. 49\,Cet was the first debris system where millimeter CO emission was detected 
\citep{zuckerman1995}. Since then several other gas species (e.g. C, C$^+$, O) have been identified 
in the disk either in absorption \citep{roberge2014} or in emission \citep{roberge2013,higuchi2017}.  
Recently the spatial distribution of the $^{12}$CO gas and dust was investigated in great 
{detail} using high angular resolution (0\farcs4) submillimeter observations with the ALMA interferometer 
\citep{hughes2017}. This study revealed a very extended, cold outer dust disk between 60 and 310\,au.
The CO emission was found to be axisymmetric and the bulk of the gas is located between  
20 and 200\,au radial distances. Since large dust grains emitting at millimeter wavelengths are not 
affected significantly by stellar radiation forces, the obtained dust map is thought to probe the 
location of the parent planetesimals. By analyzing near-infrared coronagraphic images, \citet{choquet2017} 
showed that the disk's scattered light emission is also consistent with an axisymmetric model. Small 
dust particles explored by these measurements extend from $\sim$65 to 250\,au. 

HD\,32297, a young \citep[$<$30\,Myr old,][]{kalas2005} A6-type main-sequence star, 
also harbors a well-known debris disk whose infrared excess was first identified by 
\citet{silverstone2000}. Based on Gaia DR2 astrometry the star is located  
132.3$\pm$1.0\,pc away \citep{brown2018,lindegren2018,cb2018}. Similarly to 49\,Cet, adequate fitting
of the observed excess SED requires a two-temperature model. \citet{donaldson2013} 
 derived dust temperatures of 240 and 83\,K for the warm and cold components. 
With its total fractional luminosity of 5.4$\times$10$^{-3}$ \citep{kral2017}, 
HD\,32297 belongs to the most dust-rich known debris disks.
The disk was resolved in scattered light at optical and near-IR  
wavelengths in several studies 
\citep{schneider2005,kalas2005,debes2009,currie2012,boccaletti2012,rodigas2014,esposito2014}, 
finding that it is nearly edge-on and extends far from the star 
\citep[$\sim$1800\,au\footnote{Considering the new Gaia DR2 distance estimate
instead of the Hipparcos-based 112\,pc used by \citet{schneider2014} in their paper.  },][]{schneider2014}. 
These scattered light images also revealed a strong southwest--northeast brightness asymmetry. 
Using the JCMT radio telescope, \citet{greaves2016} have recently 
detected CO (2--1) emission towards HD\,32297. 
By analyzing their recent ALMA observations of the disk, \citet{macgregor2018}
found the CO gas to be co-located with large dust particles, whose 
spatial distribution was described by the combination of a planetesimal belt 
located between 78 and 122\,au
and a halo component that extends up to 440\,au.
CO is not the sole known gas component in the disk, HD\,32297 is one of those rare 
debris disks where far-IR emission from ionized carbon was also detected 
with the {\sl Herschel Space Observatory} \citep{donaldson2013}. 
Prior to these observations, \citet{redfield2007} identified a 
stable circumstellar absorption component of \ion{Na}{1} in the optical 
spectrum of the star.

{We derived reddening values for the two stars by fitting their optical and near-IR 
 photometry with ATLAS9 models \citep{castelli1997} using fixed $T_{\rm eff}$, $\log{g}$, and 
 metallicity taken from \citet{rebollido2018}. 
 The resulting $E(B-V)$ values are 0.024$\pm$0.009\,mag and 0.028$\pm$0.013\,mag for 
 49\,Cet and HD\,32297, respectively. Combining these data with the new Gaia DR2 distances
 we computed luminosities of 17.2$\pm$0.4\,L$_\sun$ for 49\,Cet and 8.4$\pm$0.2\,L$_\sun$ for HD\,32297.}


\section{Observations} \label{sec:obs}
49\,Cet and HD\,32297 were observed using the ALMA 7-m Array in stand-alone mode in
the framework of our Cycle\,4 project (2016.2.00200.S, PI: \'A. K\'osp\'al). We
obtained Band\,6 observations of the J = 2$-$1 rotational line of $^{12}$CO, 
$^{13}$CO, and C$^{18}$O, as well as the 1.3\,mm continuum. We used two spectral 
windows for the CO line studies. The $^{13}$CO and C$^{18}$O lines were measured 
in a single baseband with a spectral resolution of 976.6\,kHz and a bandwidth of
1875\,MHz, while $^{12}$CO line was  observed in another window with a 244.1\,kHz
spectral resolution and a bandwidth of 468.8\,MHz. Two additional  spectral windows
with a bandwidth of 1875\,MHz and centred at 217.0 and 233.5\,GHz were defined to
measure the continuum emission. 
{Observations of 49 Cet were carried out in July 2017 with nine antennas 
providing projected baseline lengths between 9 and 45\,m  (6.7--34\,k$\lambda$).
The observation block began with pointing, bandpass and flux calibrations. Three quasars 
(J0522-3627, J0211+1051, J0116-1136) were used in the pointing calibrations, while quasar 
J0522-3627 and Uranus were targeted as bandpass and flux calibrators, respectively. Then, 
observations were alternated in every $\sim$6.5 minutes between 49 Cet and the 
quasar J0132-1654, that was used as a phase calibrator.
Measurements of HD\,32297 were taken in August 2017 using two observation blocks conducted 
on subsequent days (on 18 and 19 August). Nine antennas were involved providing 
projected baseline lengths from 9 to 49\,m (6.7--37\,k$\lambda$). 
Observations were executed similarly to that of 49\,Cet, i.e. the measurement started with pointing, 
bandpass and flux calibrations followed by cycles between the target and 
a phase calibrator. J0522-3627 was used as bandpass calibrator in both blocks.
In the first measurement Uranus was targeted as a flux calibrator 
and J0522-3627, J0211+1051, and J0457+0645 were used during the pointing 
 sequences. In the second measurement the quasar J0510+1800 was the flux 
calibrator, while the pointing calibration was conducted using J0522-3627
and J0510+1800.}

\begin{deluxetable}{lcc}
\tablecaption{Observational data\label{tab:imageparams}}  
\tablecolumns{3}
\tabletypesize{\scriptsize}
\tablewidth{0pt}
\tablehead{
\colhead{Parameters} & 
\colhead{49\,Cet}  &
\colhead{HD\,32297} 
}
\startdata
\cutinhead{Continuum}
Beam size ({\arcsec}) & 6$\farcs$6$\times$5$\farcs$1 & 7$\farcs$8$\times$4$\farcs$6 \\ 
Beam PA ({\degr}) & $-$81$\fdg$6 & $+$68$\fdg$5 \\
rms (mJy~beam$^{-1}$) & 0.25 & 0.11 \\
\cutinhead{$^{12}$CO (2$-$1)}
Beam size ({\arcsec}) & 6$\farcs$4$\times$4$\farcs$9 & 7$\farcs$6$\times$4$\farcs$4  \\ 
Beam PA ({\degr}) & $-$81$\fdg$5  & $+$69$\fdg$1  \\
rms (mJy~beam$^{-1}$~chan.$^{-1}$) & 28.5 & 15.2 \\
\cutinhead{$^{13}$CO (2$-$1)}
Beam size ({\arcsec}) & 6$\farcs$8$\times$5$\farcs$2 & 7$\farcs$9$\times$4$\farcs$6  \\ 
Beam PA ({\degr}) & $-$80$\fdg$6  & $+$68$\fdg$9  \\
rms (mJy~beam$^{-1}$~chan.$^{-1}$) & 17.7 & 9.0 \\
\cutinhead{C$^{18}$O (2$-$1)}
Beam size ({\arcsec}) & 6$\farcs$8$\times$5$\farcs$2 & 8$\farcs$0$\times$4$\farcs$6  \\ 
Beam PA ({\degr}) & $-$80$\fdg$6  & $+$68$\fdg$9  \\
rms (mJy~beam$^{-1}$~chan.$^{-1}$) & 11.9 & 7.7 \\
\cutinhead{Continuum flux densities and CO integrated line fluxes\tablenotemark{a}}
$F_\nu$ at 1.33\,mm (mJy)            &  5.3$\pm$0.7{(0.4)}    &   3.4$\pm$0.4{(0.1)}   \\  
$S_{\rm ^{12}CO}$  (Jy~km~s$^{-1}$) & 3.87$\pm$0.41{(0.14)}    & 1.05$\pm$0.12{(0.05)}  \\  
$S_{\rm ^{13}CO}$  (Jy~km~s$^{-1}$) & 1.68$\pm$0.19{(0.09)}    &  0.50$\pm$0.07{(0.04)}   \\  
$S_{\rm C^{18}O}$  (Jy~km~s$^{-1}$) &  $<$0.26         &  0.27$\pm$0.05{(0.04)}   \\
\enddata
\tablenotetext{a}{The quoted uncertainties include the calibration uncertainties, the numbers 
in parentheses show the measurement errors.}
\end{deluxetable}

Both data sets were calibrated and flagged with the ALMA reduction tool  {\sl
Common Astronomy Software Applications} \citep[CASA v4.7.2;][]{mcmullin2007}  using
the pipeline script delivered with the data. For 49\,Cet, data from one of the
antennas were flagged entirely by this calibration pipeline. 
We used the CASA task \texttt{uvcontsub} to fit a first order polynomial to line-free channels 
in the {\it uv} space and subtract the continuum from the obtained CO emission. 
Then the \texttt{tclean} task was used to create spectral cubes 
of the three CO lines from the continuum subtracted calibrated visibilities 
adopting natural weighting. 
The channel width was set to 0.7\,km~s$^{-1}$ for $^{12}$CO and to 
 1.4\,km~s$^{-1}$ for $^{13}$CO and C$^{18}$O lines.
To construct the continuum images we combined data from the two dedicated continuum spectral windows 
with line free channels of the two other spectral windows. 
The continuum maps were also cleaned using natural weighting. 
The achieved sensitivities as well as the size of the resulting synthesized 
beams are summarized in Table~\ref{tab:imageparams}.

\section{Results} \label{sec:res}

\subsection{Continuum Emission} \label{contemission}

\begin{figure*}
\begin{center}
\includegraphics[angle=0,scale=.24,bb=0 18 511 400]{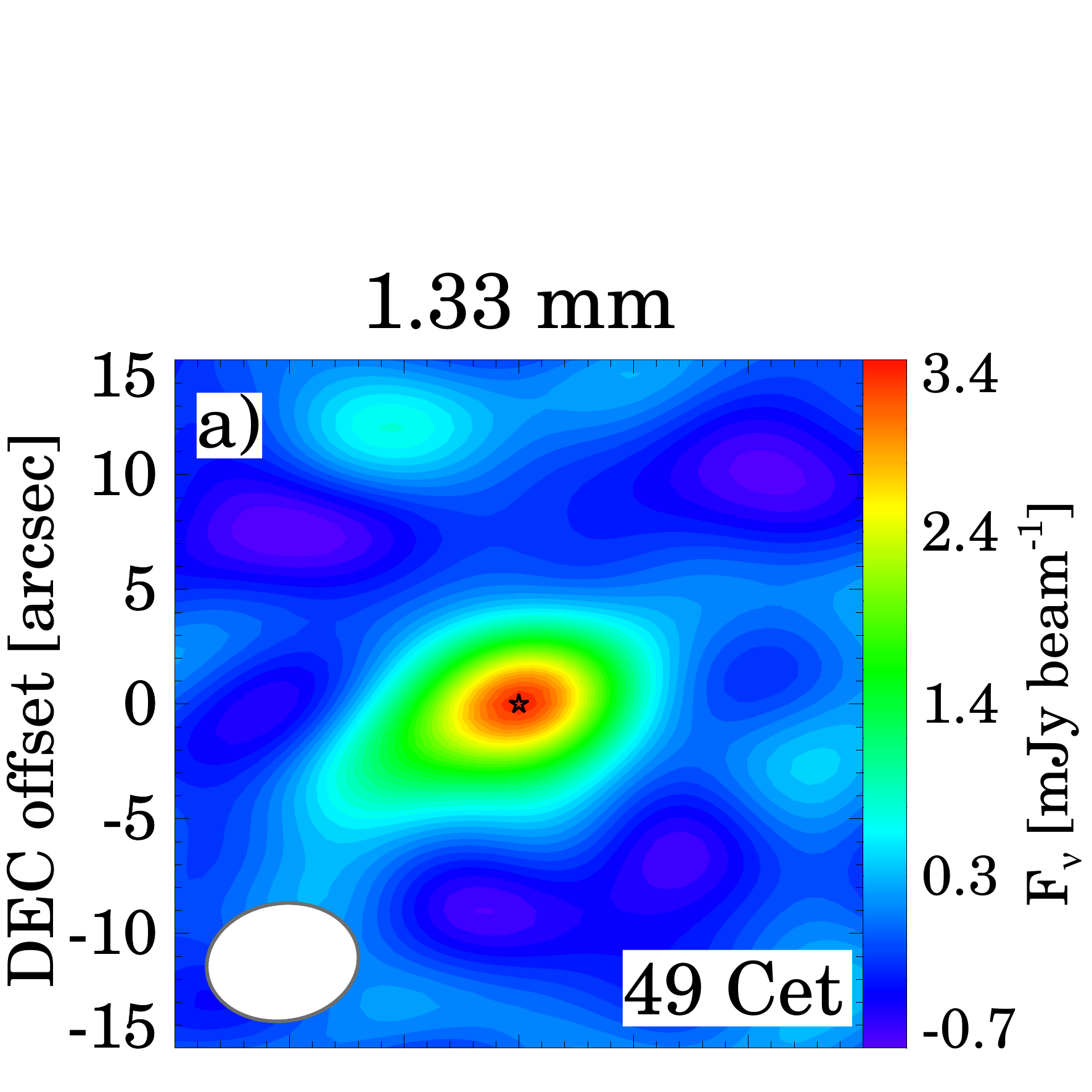}
\includegraphics[angle=0,scale=.24,bb=0 18 511 400]{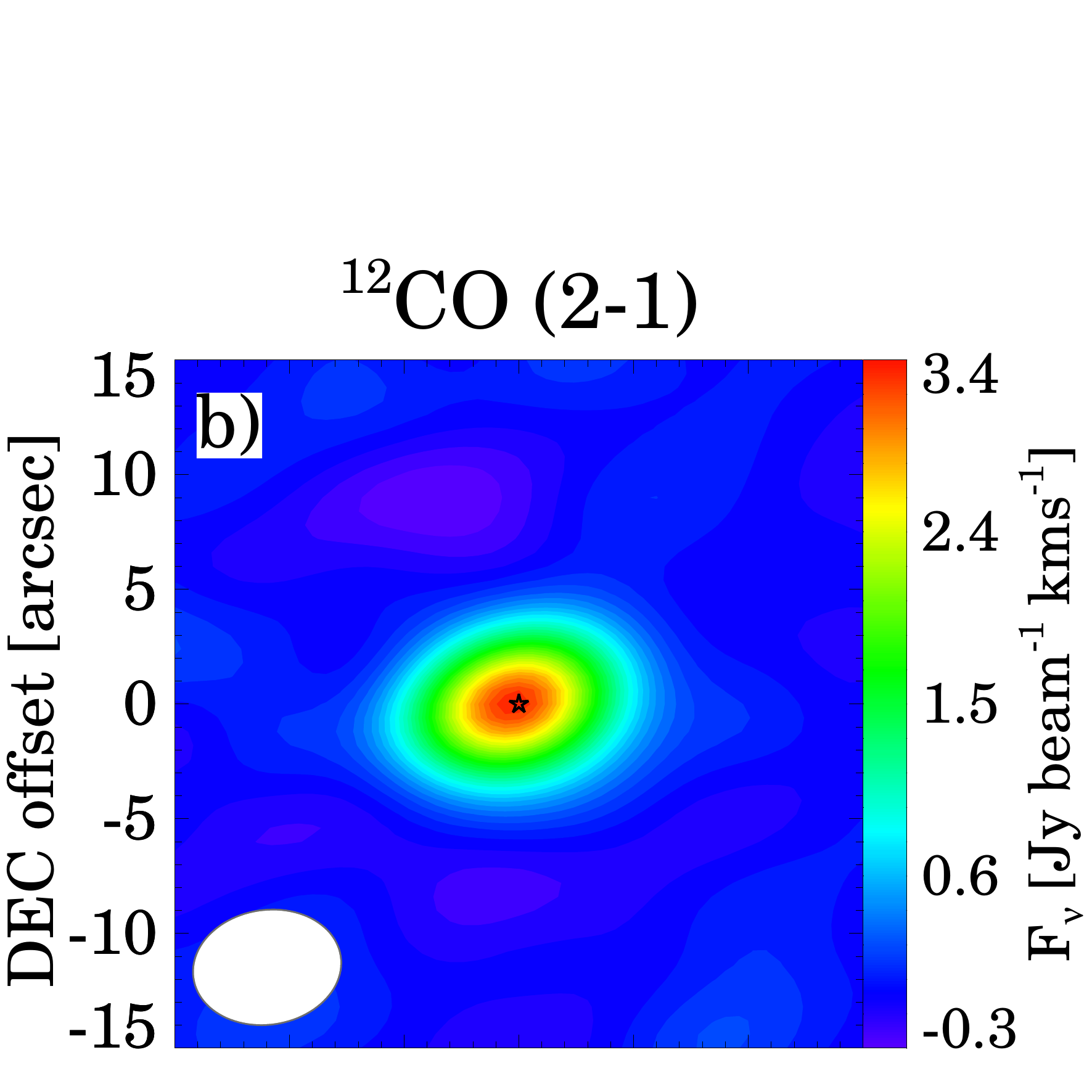}
\includegraphics[angle=0,scale=.24,bb=0 18 511 400]{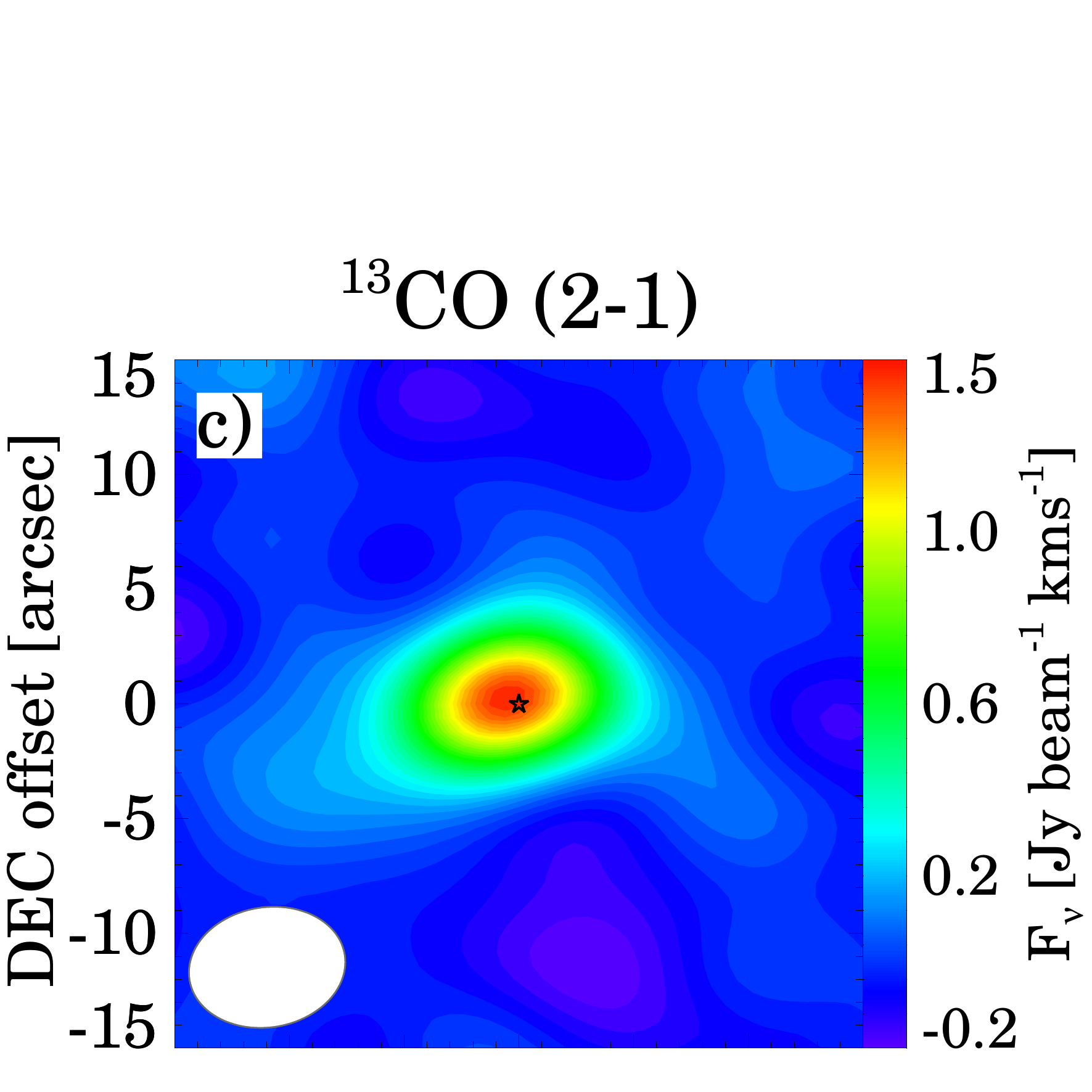}
\includegraphics[angle=0,scale=.24,bb=0 18 511 400]{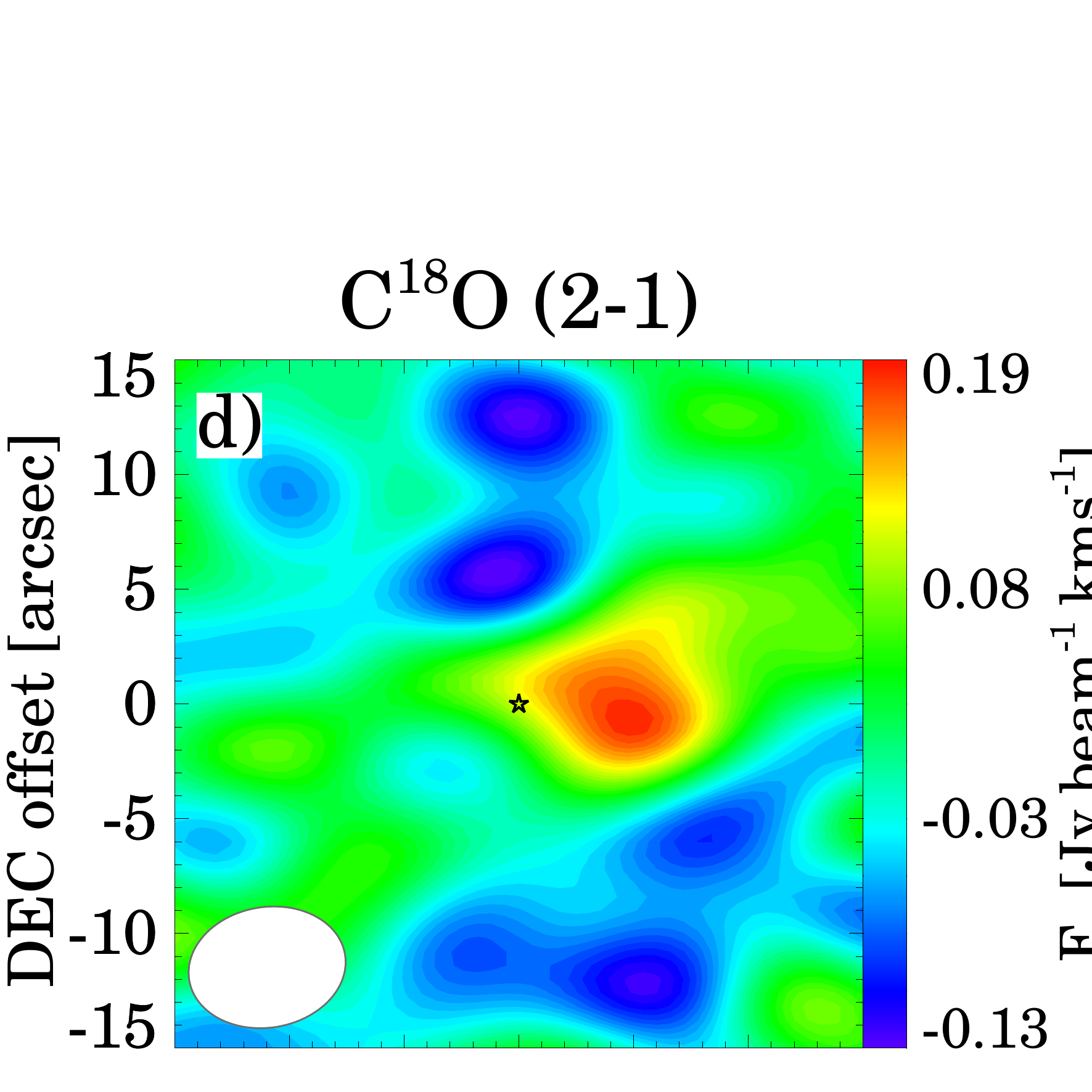}
\includegraphics[angle=0,scale=.24,bb=0 18 511 422]{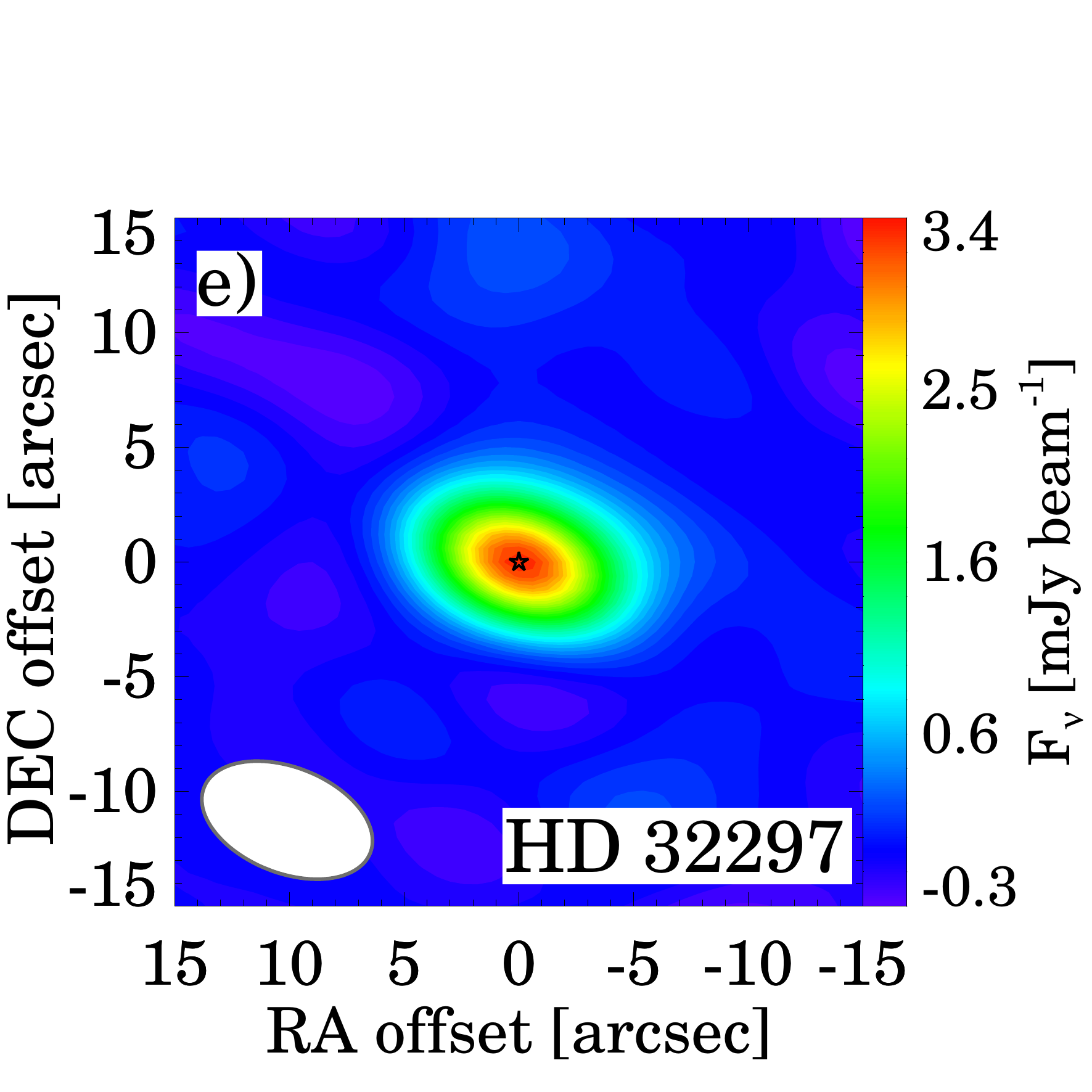}
\includegraphics[angle=0,scale=.24,bb=0 18 511 422]{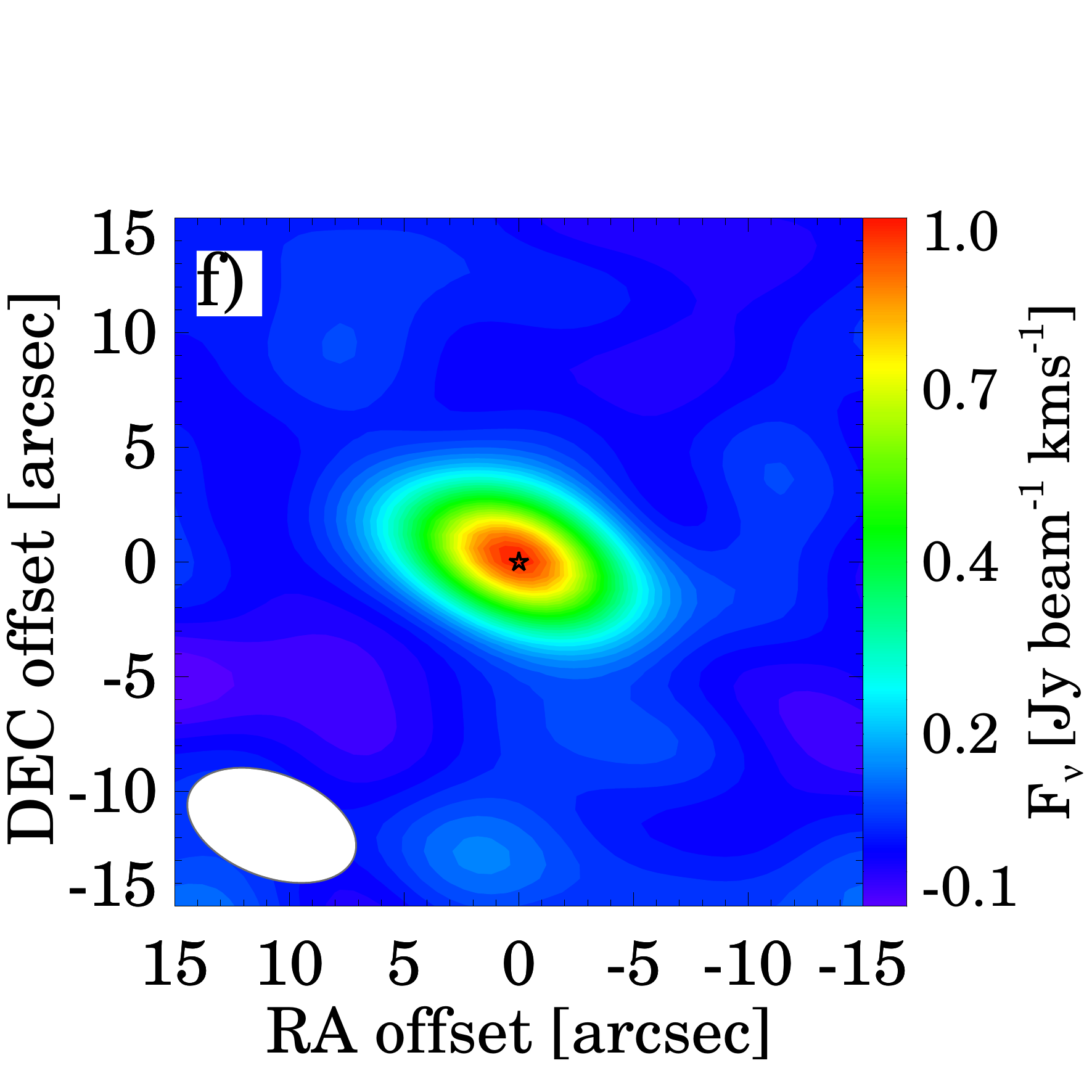}
\includegraphics[angle=0,scale=.24,bb=0 18 511 422]{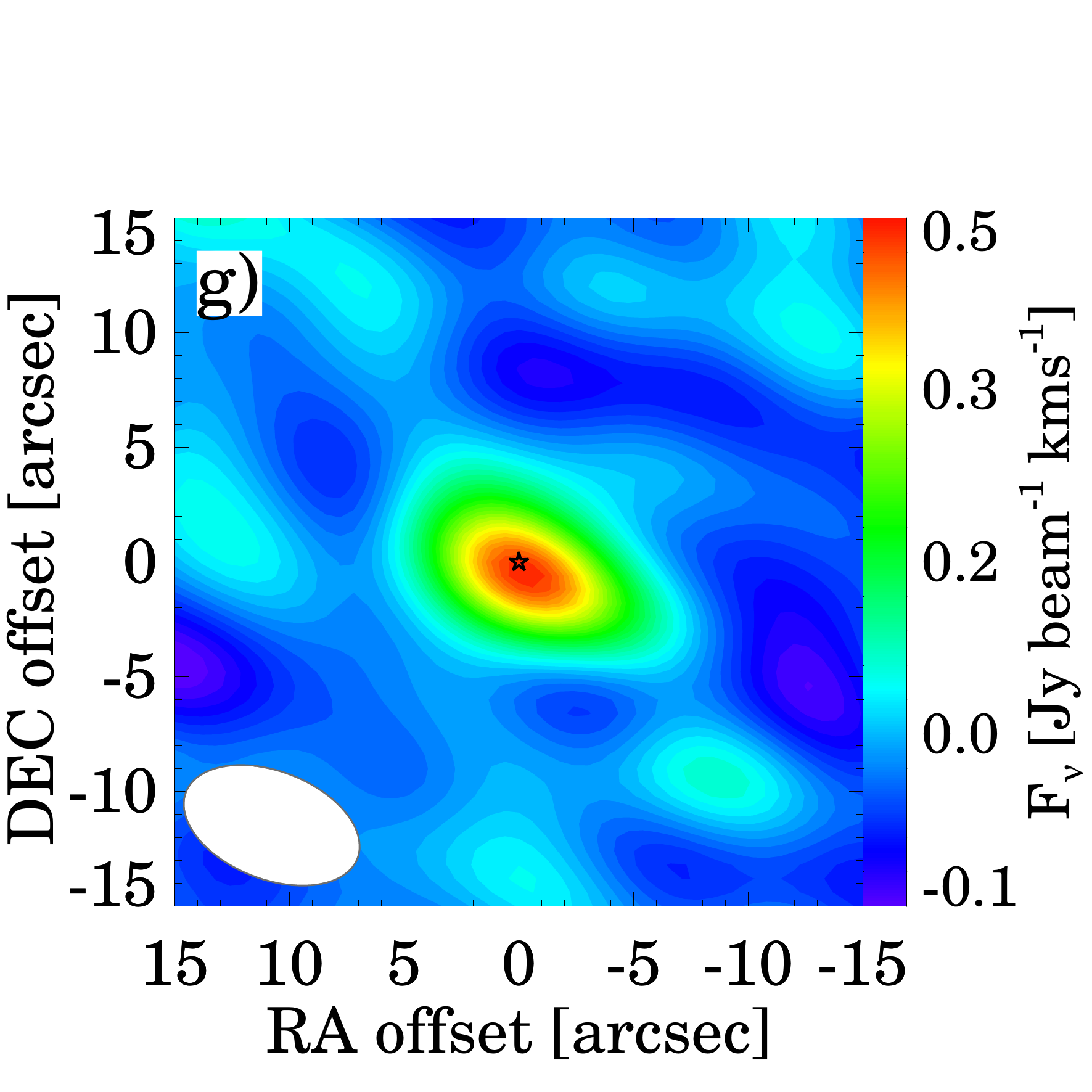}
\includegraphics[angle=0,scale=.24,bb=0 18 511 422]{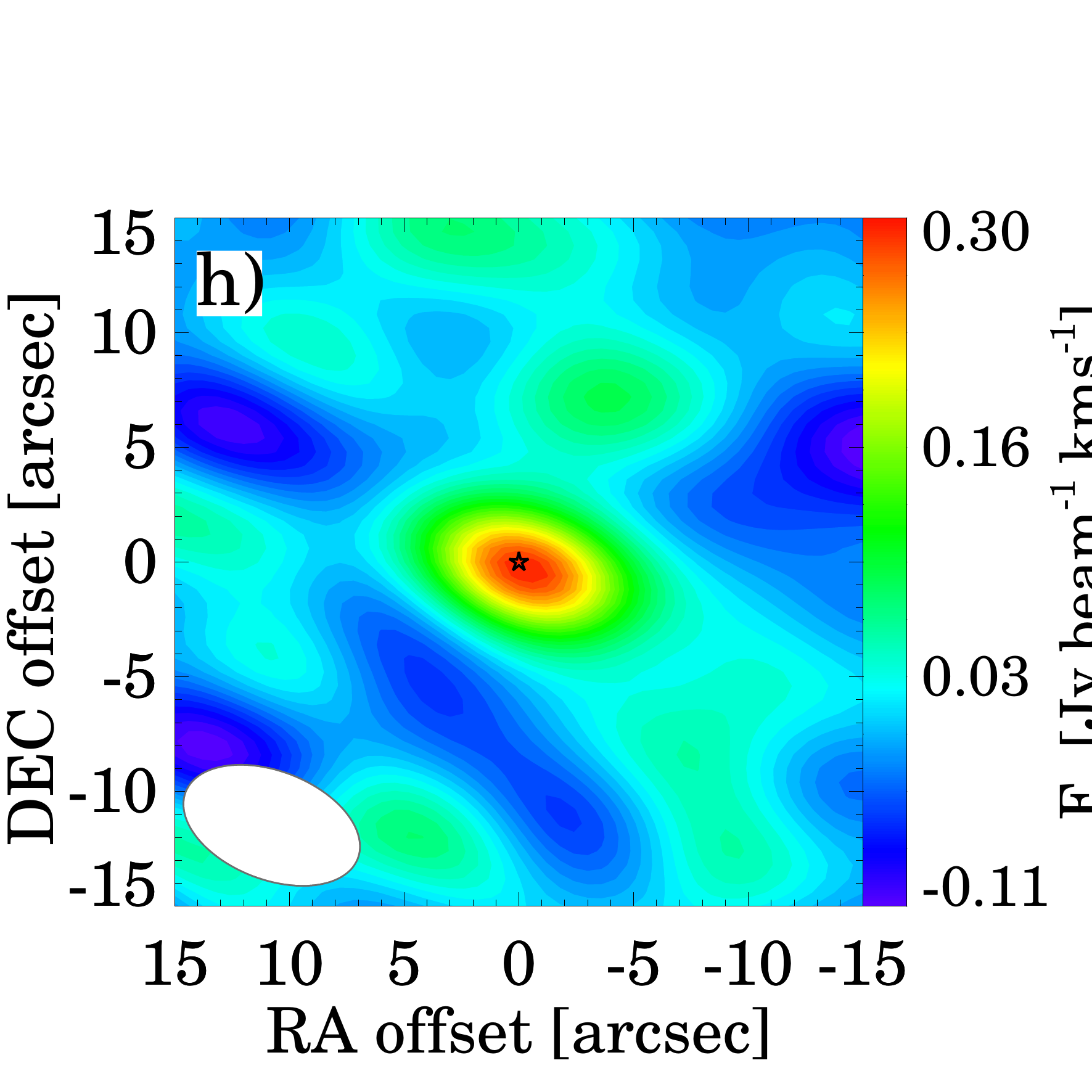}
\end{center}
\begin{center}
\includegraphics[angle=0,scale=.30,bb=0 3 563 432]{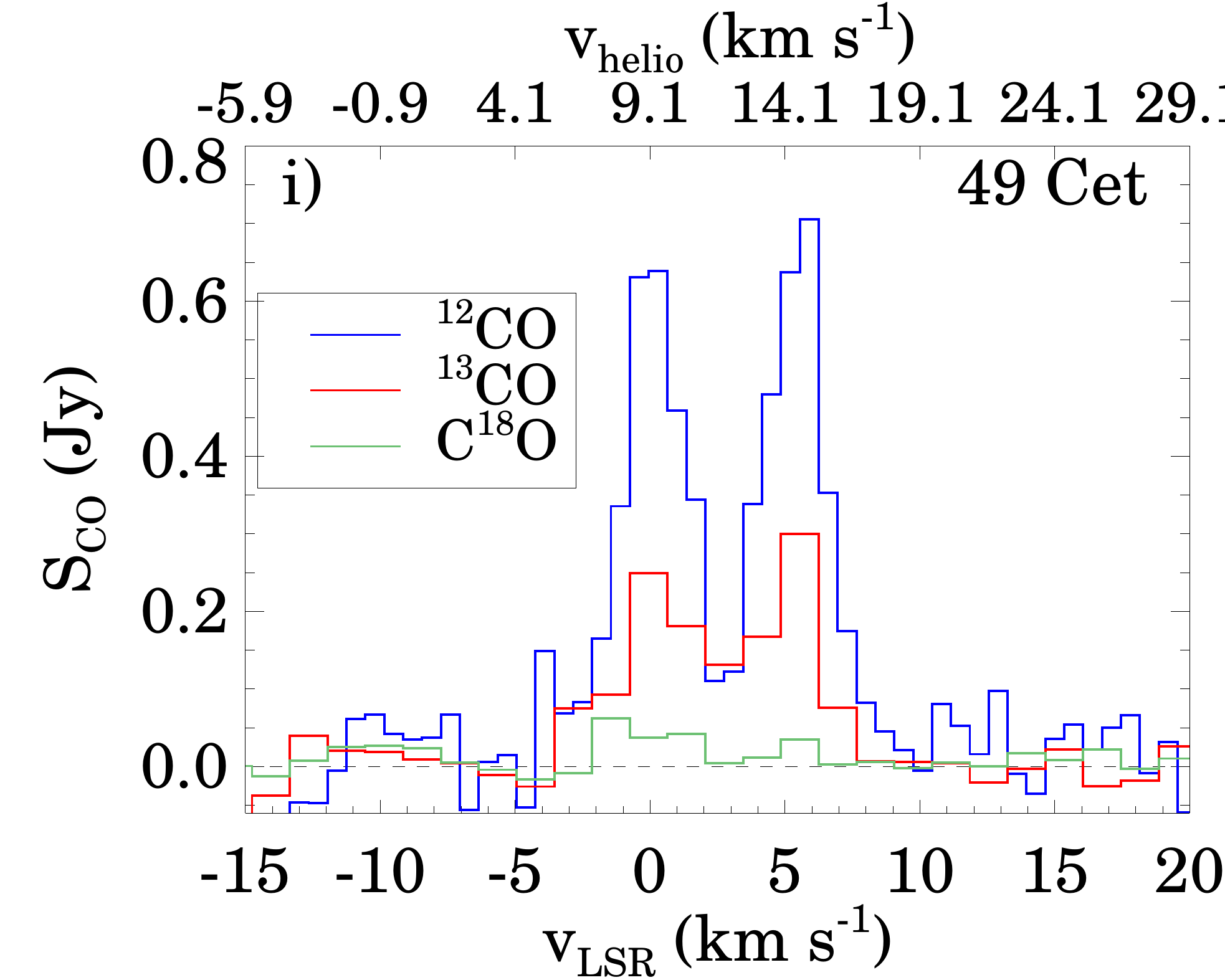}
\includegraphics[angle=0,scale=.30,bb=0 3 563 432]{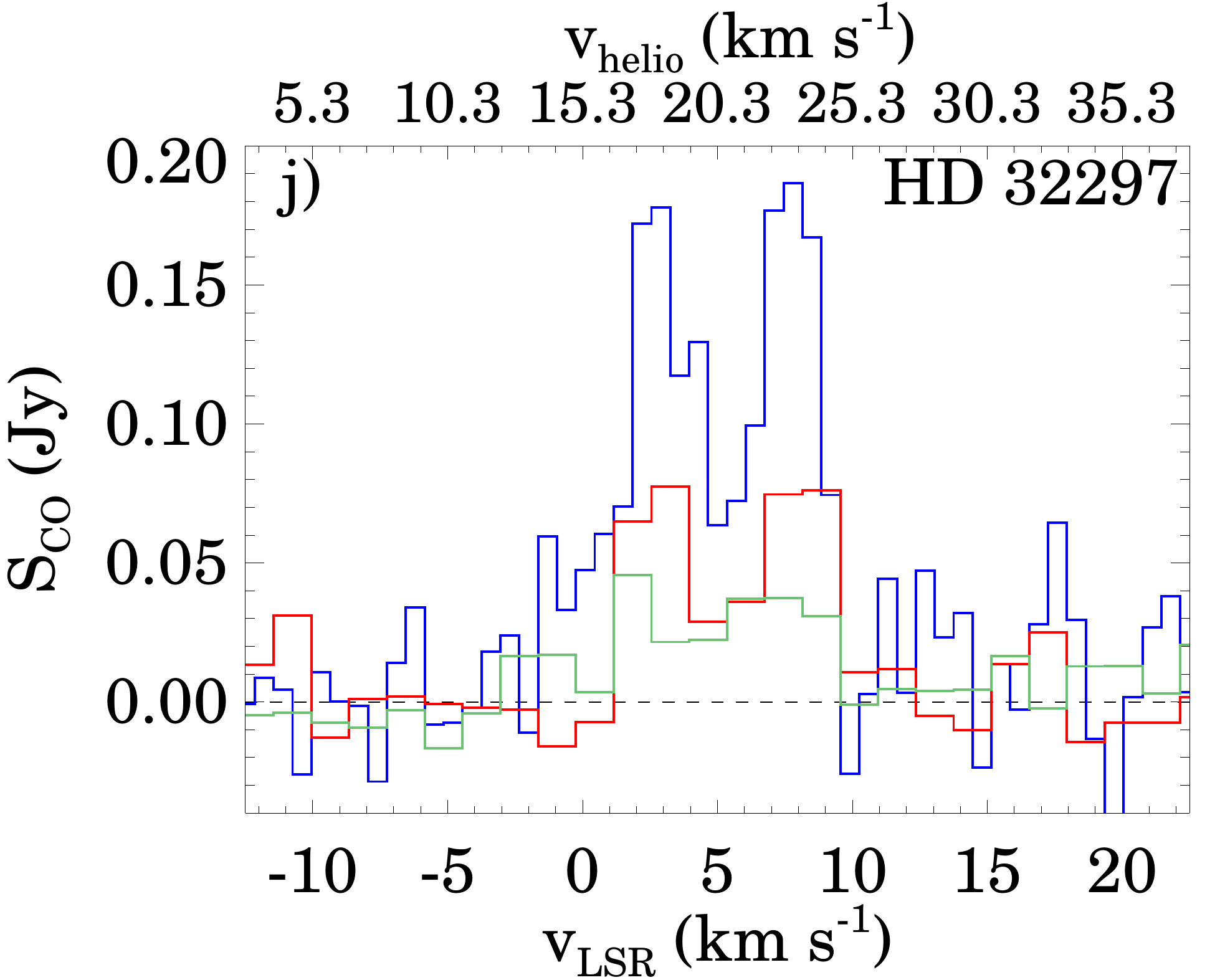}
\end{center}
\caption{Continuum and zeroth moment maps for three different CO isotopologues 
obtained with the ALMA 7\,m array in Band\,6 for 49\,Cet (first row) and 
HD\,32297 (second row) as well as the spatially 
integrated spectra of the $^{12}$CO, $^{13}$CO, C$^{18}$O emission (third row). 
\label{fig:results}}
\end{figure*}

Figures~\ref{fig:results}a and \ref{fig:results}e display the
millimeter continuum images of our targets. Both sources are clearly 
detected, at a peak signal-to-noise ratio (SNR) of 14 and 31 for 
 49\,Cet and HD\,32297, respectively.
The map of 49\,Cet (Fig.~\ref{fig:results}a) suggests 
that the disk is marginally resolved. To constrain the spatial distribution 
of emitting dust particles and to determine their total flux density at 
1.33\,mm we applied the \texttt{uvmodelfit} task in CASA that allows 
to fit a single component source model to the {\it uv} data. Channels with 
CO emission were flagged and the data weights were recomputed using the
\texttt{statwt} task.

For 49\,Cet an elliptical Gaussian model was fitted to 
the non-flagged visibility data. 
{The derived disk geometry is in good agreement with the published 
results of \citet{hughes2017} which was based on a much higher spatial resolution
ALMA data set. 
Using} \texttt{uvmodelfit} we measure a flux density of 5.3$\pm$0.7\,mJy, 
where the quoted uncertainty was computed as the quadratic sum of the measurement 
error and the typical ALMA calibration error (10\% of the signal).
By fitting the previously available (sub)millimeter photometry 
longward of 400{\micron} \citep[at 0.45, 0.5, 0.85, and 9\,mm, taken
from][]{moor2015a,macgregor2016,holland2017,hughes2017} with a straight 
line in the log-log space we can infer a flux density of 4.7\,mJy at 1.33\,mm. Our measurement 
is consistent with this prediction.

Utilizing an elliptical Gaussian model we found the disk around HD\,32297 
to be unresolved. Therefore in this case we applied a point source model 
in the visibility domain fitting, resulting in a total flux density of 3.4$\pm$0.4\,mJy. 
The obtained center of the fitted point source is consistent with the stellar 
position. Previous millimeter continuum observations of HD\,32297 with IRAM/MAMBO and with 
SMA at 1.2 and 1.3\,mm yielded, in good agreement with our result, flux densities 
of 3.14$\pm$0.82\,mJy and 3.10$\pm$0.74\,mJy \citep{meeus2012}. 
By analyzing their recent 1.3\,mm high angular resolution ALMA observations of HD\,32297, 
\citet{macgregor2018} derived a total (as a combination of a planetesimal belt and a halo component) 
flux density of 3.7$\pm$0.3\,mJy. {This result matches well the flux derived from 
our low resolution observation, which indicates that no extended emission was filtered out by ALMA in the 
earlier much higher resolution measurement.}

Assuming isothermal, optically thin dust emission and using our obtained 
millimeter flux densities, we estimated the dust mass of the disks as {\citep{hildebrand1983}}
\begin{equation}
M_{d} = \frac{{F_{\nu}} d^2}{ B_{\nu}(T_{\rm dust}) \kappa_{\nu}}, \label{eq:mdust}
\end{equation}
where $d$ is the distance, $B_{\nu} (T_{\rm dust})$ is the Planck
function at 1.33\,mm for a characteristic dust temperature $T_{\rm
  dust}$, and $\kappa_{\nu}$ is the dust opacity at the given frequency. 
{Analysis of the SEDs implied that both disks have two components 
\citep{donaldson2013,holland2017}. The millimeter emission predominantly arises 
from the colder outer belts \citep{hughes2017,macgregor2018}. For HD\,32297, \citet{macgregor2018} 
derived a characteristic temperature of 47\,K for the emitting large grains in the outer belt. 
49\,Cet harbors a very broad outer disk whose surface density peaks around 100\,au
and decreases towards larger radii \citep{hughes2017}. Therefore to estimate a characteristic temperature for 
large particles we took the equilibrium temperature of blackbody grains located 
at a radial distance of 100~au, and obtained a $T_{\rm dust}$ of 56\,K.
}
Following {\citet{beckwith1990}}, we adopted 
a value of 2.3\,cm$^2$\,g$^{-1}$ for the dust opacity. As a result we obtained dust masses of 
{0.15$\pm$0.02\,M$_\oplus$} and {0.64$\pm$0.07\,M$_\oplus$} for 49\,Cet and HD\,32297, respectively. 
The quoted uncertainties consider the errors of the measured flux density, the distance of the object, 
and of the characteristic dust temperatures, but do not account for the probable factor of $\sim$2--3 
uncertainty associated with the dust opacity \citep{miyake1993,ossenkopf1994}.

\subsection{CO Line Emission} \label{coemission}
By inspecting our CO data cubes in addition to detection of $^{12}$CO emission 
at high peak SNR ($>$12) we discovered $^{13}$CO emission at peak   
SNRs of $>$9 around the position and 
radial velocity of both stars.
For HD\,32297 even the C$^{18}$O line was clearly detected. We started the analysis with the
$^{12}$CO and $^{13}$CO data.   We applied different-sized elliptical apertures with an
axis ratio and a position angle identical to the appropriate beam to extract the line
spectra from the data cubes and to determine the integrated line flux by summing all
consecutive channels that show significant line emission, i.e. where the peak SNR is 
higher than 3 (within the aperture).
By examining the obtained
integrated line fluxes as a function of the applied apertures then  we defined the
minimum aperture size that includes all CO emission associated to the disk. 
{We obtained identical aperture sizes for $^{12}$CO and $^{13}$CO data at both disks.} 
In the case
of C$^{18}$O data the integration was performed with the same aperture radius and  over
the same velocity range as established for the more abundant 
species.
To estimate the statistical error of the derived line fluxes we used the line-free regions of 
the spectra, the final uncertainty was computed by adding a 10\% flux calibration 
error to this.  
Zeroth moment maps and the acquired spectra for the examined transitions
are displayed in Figure~\ref{fig:results}, the obtained integrated line fluxes 
with their uncertainties are listed in Table~\ref{tab:imageparams}.  
In the case of 49\,Cet no significant 
C$^{18}$O emission was detected at the position of the star but a source appears 
close to it with an offset of 5$\farcs$7 (the positional uncertainty, 
estimated as the ratio of the beam size to achieved SNR, is $\sim$1\farcs7). 
By centering the aperture on this source 
we obtained a spectrum whose shape resembles well that of the $^{12}$CO and $^{13}$CO   
lines of 49\,Cet. Integration over the above specified velocity range results in  
an integrated line flux of 257$\pm$58\,mJy~km~s$^{-1}$. The similarity of the spectrum 
and the fact that there is no counterpart of this source in the $^{12}$CO and 
$^{13}$CO maps suggest that the observed emission comes from the disk.  
However, since we have no explanation for the observed large offset, 
the integrated line flux is considered as an upper limit 
in the following analysis.        

As Figure~\ref{fig:results} (i and j) demonstrates, the obtained CO spectra show typical 
double-peaked profiles indicating that the observed emission arises from 
rotating gas. Using the ALMA 12m array \citet{hughes2017} obtained a 
high-SNR $^{12}$CO (3--2) line spectrum for 49\,Cet (see their fig.~3). 
The shape and width of our $^{12}$CO (2--1) line are consistent with 
those of that 3--2 spectrum. Previous observation of the 2--1 line 
with the Submillimeter Array (SMA) resulted in a line flux of 
2.0$\pm$0.3\,Jy~km~s$^{-1}$ \citep{hughes2008}, which is about two times lower than our 
value. 

The first detection of CO emission towards HD\,32297 was made by the 15\,m 
James Clerk Maxwell Telescope (JCMT) radio telescope \citep{greaves2016}. 
The obtained $^{12}$CO (2--1) line profile \citep[fig.~3 in][]{greaves2016} is 
$\sim$18\,km~s$^{-1}$ wide. 
The line in our observation is only about half as broad
and mostly corresponds to the blue wing of the JCMT spectrum in terms of its 
velocity range. 
\citet{greaves2016} used two different baseline fitting models and derived
peak fluxes of $\sim${320\,mJy} and $\sim${440\,mJy}, i.e. $>$1.8 
times higher than in our case. As a consequence of the mentioned differences their derived 
integrated line fluxes (2.7\,Jy~km~s$^{-1}$ or 5.1\,Jy~km~s$^{-1}$, depending 
on the applied baseline model) are also significantly higher than ours. 
Recently \citet{macgregor2018} presented a $^{12}$CO (2--1) 
observation for HD\,32297 with ALMA. The velocity interval 
of the CO emission (as shown in their position-velocity diagram, their fig.~3, left) as 
well as the derived 
integrated line flux of 1.02\,Jy~km~s$^{-1}$ 
are consistent with our results. {Therefore the somewhat deviating line observation  
of \citet{greaves2016}, derived from 
a lower S/N measurement, will not be taken into account in the further 
analysis.}     

Using the obtained line fluxes, we derived line flux ratios of 
$S_{\rm ^{12}CO} / S_{\rm ^{13}CO} = 2.3\pm0.2$ 
and $S_{\rm ^{13}CO} / S_{\rm C^{18}O} > 6.5$ for 49\,Cet and 
$S_{\rm ^{12}CO} / S_{\rm ^{13}CO} = 2.1\pm0.2$ 
and $S_{\rm ^{13}CO} / S_{\rm C^{18}O} = 1.8\pm0.3$ for HD\,32297. Taking isotope ratios of 
[$^{12}$C]/[$^{13}$C]=77 and [$^{16}$O]/[$^{18}$O]=560, typical
in the local interstellar matter \citep{wilson1994}, and assuming that all three isotopologues are optically 
thin and have identical excitation temperatures between 10 and 100\,K,  one would expect line 
ratios of $S_{\rm ^{12}CO} / S_{\rm ^{13}CO} \sim 85-91$ and $S_{\rm ^{13}CO} / S_{\rm ^{18}CO} \sim 7.3$. 
Thus the measured values imply highly optically thick $^{12}$CO emission in both disks, while 
the $^{13}$CO line is probably optically thick in HD\,32297 and optically thin 
in 49\,Cet.
 
We used the optically thin line observations to estimate the CO gas mass. As a first step 
we derived the mass of the specific $^{y}$C$^{z}$O isotopologue as:
\begin{equation}  
M_{^{y}C^{z}O} = {4 \pi m d^2} \frac{S_{21}}{x_2 h \nu_{21} A_{21}}, \label{eq:co}
\end{equation}
where $m$ is the mass of the given molecule, $d$ is the distance of the target, $h$   
is the Planck constant. $S_{21}$ is the integrated line flux, $A_{21}$ and $\nu_{21}$
are the Einstein coefficient and the rest frequency of the specific rotational transition, 
while $x_2$ is the fractional population of the upper level. 

We assumed that local 
thermodynamical equilibrium (LTE) holds and utilized the Boltzmann equation
to compute the fractional populations. 
{Our observations do not allow to derive reliable gas temperatures 
for the targeted disks.  
Gas temperature estimates in CO-rich debris 
disks are uncertain, and previous studies typically resulted in low values. Based on our CO observations, 
we derived $\lesssim$10K for the disk of HD~21997 \citep{kospal2013}. Using the 
measured peak flux value of the high angular resolution $^{12}$CO (3--2) map of 49~Cet 
\citep{hughes2017} we could derive a maximum brightness temperature of $\sim$26\,K for 
that disk. For HD\,141569, a system that resembles the most CO-rich debris disks in terms of 
dust/CO mass (Sect.~\ref{sec:cocontent}), \citet{flaherty2016} found a gas temperature of 27$^{+11}_{-4}$\,K at a radial 
distance of 150\,au by modelling CO observations. In all these cases the derived 
temperatures are lower than the characteristic dust temperatures of the given systems.}  

{Considering these results, for the excitation temperature we adopted 20\,K 
for both our disks.} 
Based on Eq.~\ref{eq:co} we then obtained 
$M_{\rm ^{13}C^{16}O}$ = (1.5$\pm$0.2)$\times$10$^{-4}$\,M$_\oplus$
for 49\,Cet
and $M_{\rm ^{12}C^{18}O}$ = (1.4$\pm$0.3)$\times$10$^{-4}$\,M$_\oplus$ for HD\,32297. 
The uncertainties 
account only for errors in the integrated line fluxes and distances of the disks.  
Different gas temperatures and/or non-LTE (NLTE) conditions would result in different 
$x_2$ values and thereby gas masses. Since with the adopted temperatures the fractional population 
is close to its maximum in LTE, these deviations can typically lead to higher gas 
masses (lower $x_2$). {We note that if LTE holds, for gas temperatures between 
7 and 74\,K the gas mass estimates would not exceed our derived CO masses by more than 
a factor of two.}   

To estimate the total $^{12}$CO gas quantity, the obtained CO isotopologue 
masses were multiplied by the corresponding isotope ratios  
and took into account the difference in the molecular masses. 
These yielded CO gas mass of 
$M_{\rm ^{12}C^{16}O}$ = (1.11$\pm$0.13)$\times$10$^{-2}$\,M$_\oplus$
and 
(7.4$\pm$1.3)$\times$10$^{-2}$\,M$_\oplus$ for 49\,Cet and HD\,32297, respectively. 
Different physical/chemical 
mechanisms, perhaps most importantly the isotope selective photodissociation can alter 
the abundance of the isotopologues, resulting in isotope ratios substantially different 
from the ones typical in the local interstellar medium \citep[e.g.,][]{visser2009}. 
This can introduce serious uncertainties in the estimates of the total CO gas 
mass \citep[see Sect.~\ref{sec:shielding} and e.g.,][]{miotello2014}.


\section{Discussion} \label{sec:discussion}

\subsection{Young debris disks with large CO content}  \label{sec:cocontent}

\begin{figure}
\begin{center}
\includegraphics[angle=0,scale=.48]{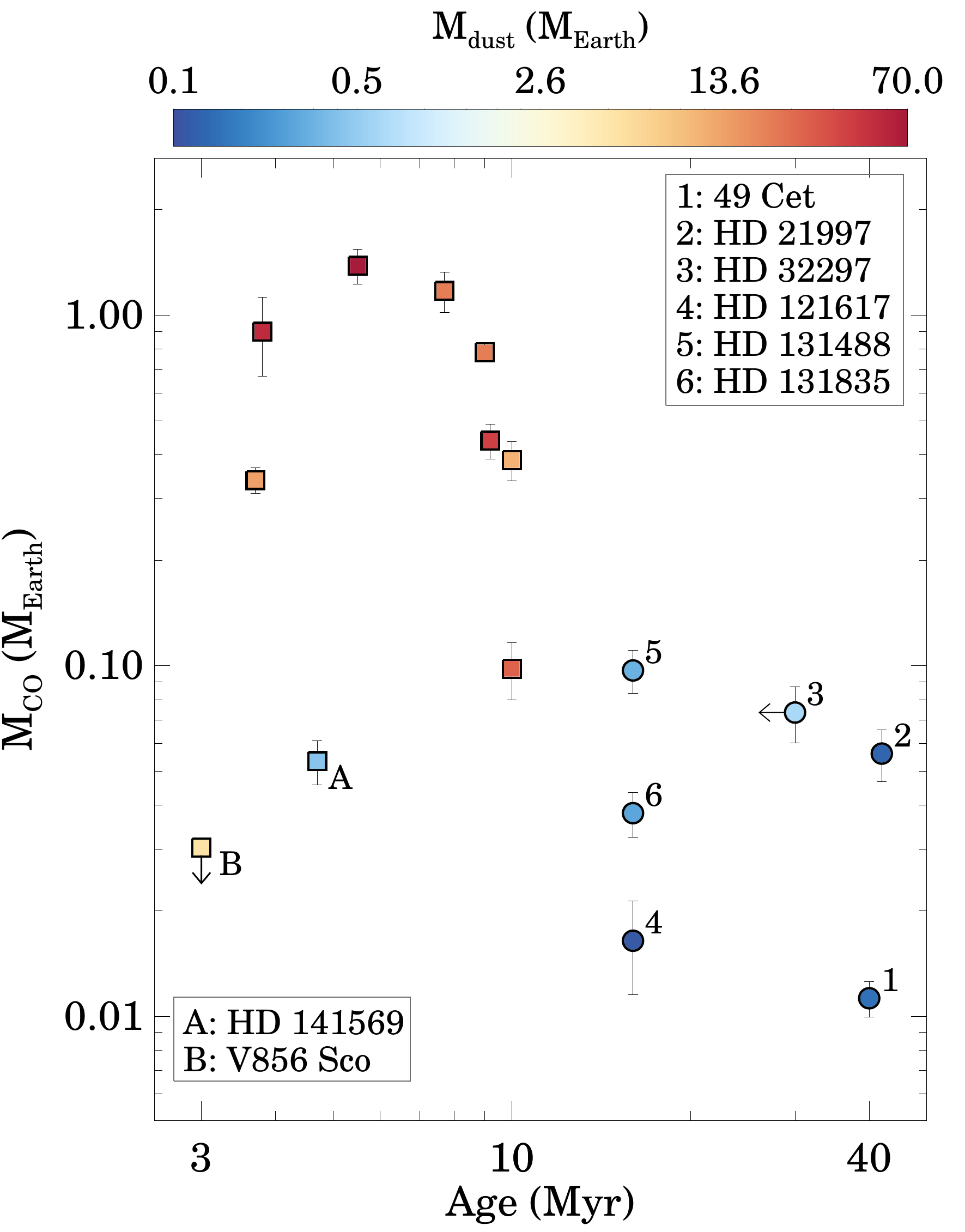}
\end{center}
\caption{{CO masses of circumstellar disks as a function of age. Dust masses are 
also shown by the colors of the symbols. 
Debris disks are plotted by circles, while disks around Herbig Ae stars are 
displayed by squares.
Data for 49\,Cet and
HD\,32297 are taken from Sect.~\ref{contemission}-\ref{coemission}. For the other four CO-rich debris 
disks we used data from \citet{moor2017}, but considering their new Gaia DR2 based 
distances from \citet{cb2018}.
To estimate the CO masses of protoplanetary disks we 
followed the outline described in Sect.~\ref{coemission}. Apart from the case of HD\,141569, 
where a $^{13}$CO observation was used \citep{pericaud2016b}, all our calculations are based
on C$^{18}$O line fluxes that are taken from the literature 
\citep[][]{ansdell2016,hales2018,favre2013,kastner2018,rosenfeld2013,fedele2017,vanderplas2019} 
or derived by processing data from the ALMA archive. Dust masses are derived from 
(sub)millimeter continuum measurements of the targets 
\citep{andrews2012,ansdell2016,chapillon2008,hales2018,henning1994,isella2007,kastner2018,meeus2012,
raman2006,sylvester1996}, assuming optically thin emission. For Herbig\,Ae stars we adopted 
gas and dust temperatures of 50\,K in these calculations.} 
\label{fig:mco}}
\end{figure}

Our observations of rarer CO isotopologues in debris disks around 49\,Cet and HD\,32297 
implied that their $^{12}$CO emission is optically thick and their total CO mass is higher 
than 0.01\,$M_\oplus$. These two objects do not stand alone with this characteristic, 
previous measurements revealed similarly CO-rich debris disks around young A-type stars 
HD\,21997, HD\,121617, HD\,131488, and HD\,131835. The estimated CO content of these 
systems is on average three orders of magnitudes higher than that of the gaseous debris 
disk of $\beta$\,Pic \citep[$\sim$3.6$\times$10$^{-5}$\,$M_\oplus$,][]{matra2018a} and is 
more similar to those of protoplanetary disks. {To explore the latter point further, in
Figure~\ref{fig:mco} we plotted the CO masses of these six debris disks (circles) along with 
disks around ten nearby ($\lesssim$160\,pc) Herbig\,Ae stars (squares) as a function of age. 
Dust masses of the disks are also shown by colors of the symbols.
These Herbig\,Ae disks can be considered as the precursors of young debris disks 
with A-type host stars. While their average dust mass is about two 
orders of magnitude higher than that of our debris disks, in terms of their CO gas quantity 
 the difference is only one order of magnitude on average and there are some 
less massive Herbig\,Ae disks (e.g. V856\,Sco) with CO content comparable to 
that of the debris sample. The disk around the well known pre-main-sequence star, HD\,141569, 
resembles CO-rich debris disks both in terms of dust and CO masses.}

\subsection{Shielding of CO gas in 49\,Cet and HD\,32297}  \label{sec:shielding}

Does the protoplanetary-like CO content and the high CO-to-dust mass ratios inevitably mean 
that 49\,Cet and HD\,32297 (as well as the other four debris disks) harbor 
primordial gas? An ultimate way to answer this question would be to measure 
the H$_2$ content of these disks. If the gas, similarly to protoplanetary disks, 
is predominantly primordial then hydrogen molecules are the primary gaseous species. 
On the other hand, if the gas has a secondary origin then the abundance of H$_2$ molecules 
in the gas mixture is expected to be low \citep{kral2017,kral2018}, and the gas is rather composed of 
different molecules released from the icy bodies and their photodissociation products. 
This also means that the observed comparable CO masses do not neccessarily indicate 
similar total gas masses.
Unfortunately, the detection of molecular hydrogen is notoriously difficult. Therefore here we use 
a different strategy and investigate whether the observed CO gas quantity of 
49\,Cet and HD\,32297 could be explained within the framework of a secondary gas production 
scenario \citep[see][]{kral2018}.

High angular resolution millimeter continuum observations of 49\,Cet and HD\,32297 
\citep{hughes2017,macgregor2018} implied that they -- similarly to the other four CO-debris 
disks mentioned above -- harbor massive, cold ($<$110\,K) dust belts at radial distances larger 
than $\sim$30\,au with which the observed CO gas is at least partly co-located. Supposing that 
the parent planetesimals are icy -- which is permitted by the typical temperatures in the disks of 49\,Cet and 
HD\,32297 
-- mutual collisions between these bodies produce not only smaller and smaller particles but 
can release gas as well \citep{moor2011,zuckerman2012,kral2017}. Based on volatiles detected in 
spectroscopic surveys of Solar System comets \citep{mumma2011}, water, carbon-monoxide, and 
carbon-dioxide may be the most important products of this process. Due to stellar and 
interstellar UV photons, then the released gas molecules are photodissociated. 
According to \citet{kral2017}, assuming a steady state balance of gas production and loss,
the CO mass of a disk can be estimated as  
\begin{equation}
M_{\rm CO} = \dot{M}_{\rm CO} t_{\rm ph} \epsilon_{\rm CO} = \dot{M}_{\rm loss} \gamma t_{\rm ph} 
\epsilon_{\rm CO},  \label{eq:coprod}
\end{equation}
where $\dot{M}_{\rm loss}$ is the mass loss rate in the collisional cascade, $\gamma$ is the 
CO+CO$_2$ ice mass fraction of planetesimals, $t_{\rm ph}$ is 
the photodissociation timescale of unshielded CO gas, while $\epsilon_{\rm CO}$ is a factor
that shows how well shielded the molecules are. Here we consider that photodissociation of CO$_2$ molecules 
also contribute to the observed CO gas. If we know $t_{\rm ph}$, $\gamma$, and 
$\dot{M}_{\rm loss}$ then the minimum necessary shielding factor in the 49\,Cet and HD\,32297 
systems can be estimated. 
As a lower threshold for the UV flux, we consider only the influence of interstellar UV photons 
and following \citet{visser2009} we set a dissociation timescale of $\sim$120\,yrs for unshielded CO molecules. 
In Solar System comets the ice mass fraction is not higher than 0.27 \citep{mumma2011,matra2017a}, 
thus we adopt this value for the $\gamma$ parameter. 
Assuming a steady-state collisional cascade in the planetesimal belt, we estimated the mass loss rate 
using the formula proposed by \citet[][see their eq.~21]{matra2017b}: 
\begin{equation}
\dot{M}_{\rm loss} ({M_\oplus}/{\rm Myr}) = 1200 \bigg( \frac{R}{\rm au} \bigg)^{1.5}
                     \bigg(\frac{\Delta R}{\rm au} \bigg)^{-1} f_d^2 
		     \bigg(\frac{L_*}{L_\odot}\bigg) 
		     \bigg(\frac{M_*}{M_\odot} \bigg)^{-0.5} 
\label{eq:dustprod}.
\end{equation}
Based on \citet{macgregor2018} for the outer belt of HD\,32297 we adopted a radius ($R$) and width ($\Delta R$) 
of 100\,au and 
40\,au, respectively.  In the case of the very broad disk of 49\,Cet we used the double power-law surface density 
profile 
model proposed by \citet{hughes2017} and the inner (71\,au) and outer radii (153\,au) of the disk was set at 
half of the surface density 
maximum. This resulted in $R=112$\,au and $\Delta R = 82$\,au. 
Fractional dust luminosities and stellar parameters needed for this
calculation were taken from Sect.~\ref{sec:targets} and from the literature \citep{hughes2017,macgregor2018}.
Using Eq.~\ref{eq:dustprod} then we obtained mass loss rates of 0.1 and 5.2\,M$_\oplus$~Myr$^{-1}$ for 49\,Cet and 
 HD\,32297, respectively.
Applying these values in Eq.~\ref{eq:coprod} yielded shielding factors of 
$\epsilon_{\rm CO}\sim$3600 and $\sim$400 for 49\,Cet and HD\,32297, respectively. 
Since the ice mass fraction is likely lower than 0.27 and -- especially in the case 
of 49\,Cet -- stellar UV photons can affect the dissociation of CO molecules 
in the inner part of the disk these factors should be considered as lower limits.  
We note that the lower $\epsilon_{\rm CO}$ value of HD\,32297 is due to its 
prominently high fractional luminosity (the dust production rate being proportional to $f_d^2$). 

\begin{figure}
\begin{center}
\includegraphics[angle=0,scale=.50]{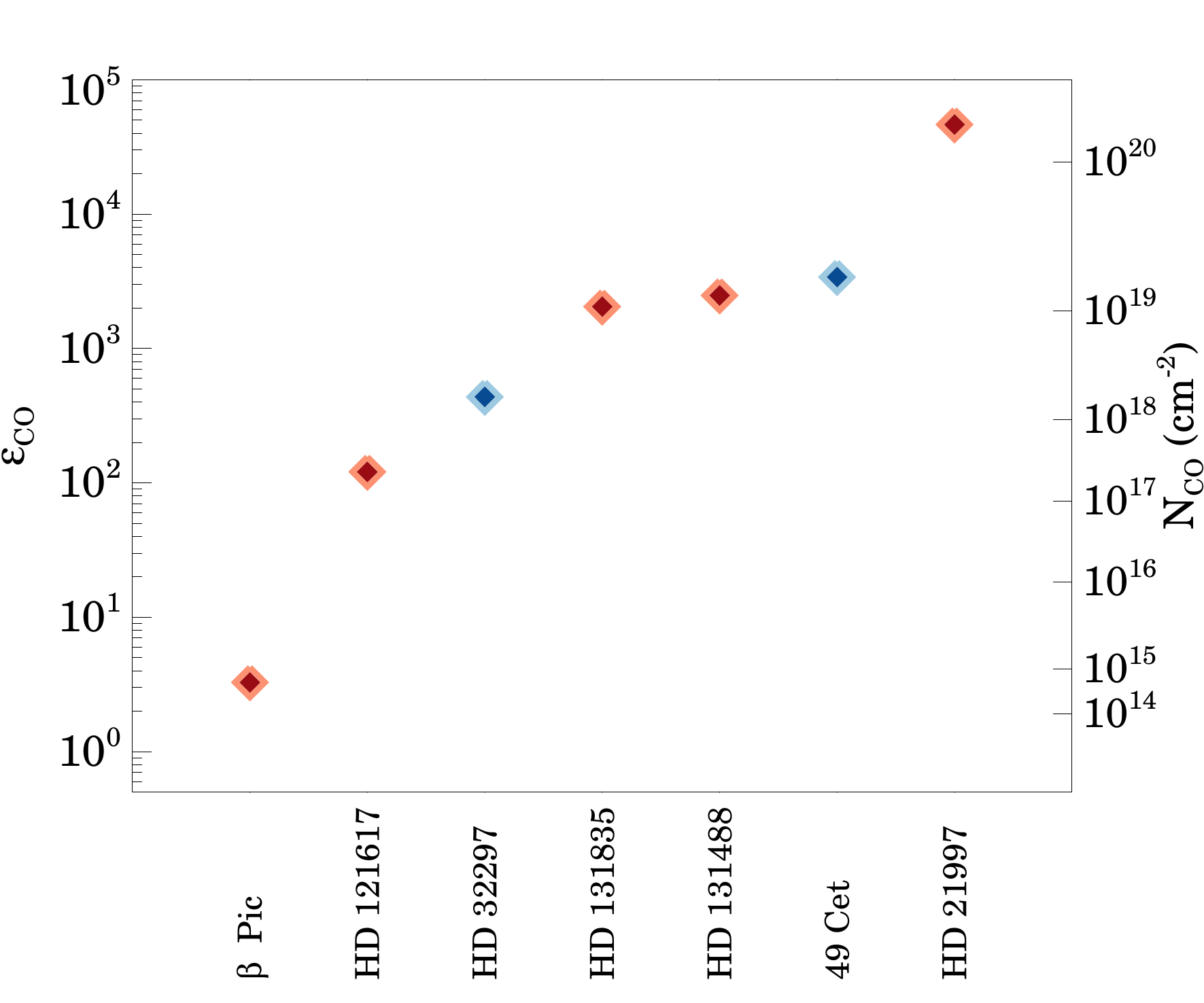}
\end{center}
\caption{Shielding factors necessary to explain the CO content of 
the six CO-rich debris disks and the disk of $\beta$\,Pic (\ref{sec:shielding}). 49\,Cet and 
HD\,32297, the targets of our study, are plotted with blue symbols. 
CO column densities corresponding to the given shielding factors are also drawn in the right-hand side. 
These column densities were computed by interpolating in table~5 of \citet{visser2009}. 
We note that in the case of HD\,21997, where the required shielding factor is extremely high, we had to use 
extrapolation to obtain the relevant CO column density. 
\label{fig:shielding}}
\end{figure}

What processes can contribute to the shielding in these systems?
While in protoplanetary disks continuum shielding by dust can substantially reduce 
the flux of incoming UV photons, in debris disks their absorption by optically thin
dust is insignificant. Since photodissociation of CO molecules occurs primarily through 
discrete absorptions into bound electronic states, they are subject to self-shielding.
According to \citet{visser2009} at $^{12}$CO column densities higher than 
10$^{15}$\,cm$^{-2}$ self-shielding can reduce significantly the photodissociation 
rate of $^{12}$CO molecules located in deeper regions. By interpolating in their tab.\,5 we found 
that to achieve 
the above shielding factors the vertical $^{12}$CO column density in the 
disks around 49\,Cet and HD\,32297 must exceed $\sim$1.7$\times10^{19}$\,cm$^{-2}$ and 
$\sim$1.5$\times10^{18}$\,cm$^{-2}$, respectively. Though \citet{macgregor2018} did not provide 
detailed model for the distribution of CO gas in HD\,32297 they argued that it is 
co-located with dust. Assuming that the gas follows the same spatial distribution 
as the dust component and taking the total CO mass derived in Sect.~\ref{coemission}, we 
derived a maximum vertical column density of $\sim$10$^{18}$\,cm$^{-2}$ for this disk. This 
indicates that at least in the densest regions the self-shielding mechanism can play 
an important role in the long-term survival of $^{12}$CO molecules. In the case of the less CO-rich 
49\,Cet, the typical vertical CO column densities fall short of the value required 
for effective shielding. In Figure~\ref{fig:shielding} we display the necessary shielding 
factors for our two targets together with the other four high CO mass debris 
disks and the disk of $\beta$\,Pic. 
Stellar and disk parameters needed for calculations of $\dot{M}_{\rm loss}$ were taken from the 
literature \citep{lieman-sifry2016,hughes2017,kral2018,moor2017,matra2018b,macgregor2018}. The obtained $\dot{M}_{\rm loss}$ 
values range between 0.04 and 4.2\,M$_\oplus$~Myr$^{-1}$. 
The CO mass estimates are from Fig.~\ref{fig:mco}a, except for $\beta$\,Pic where it was taken from \citet{matra2018a}. 
As the plot shows, while the CO content of $\beta$\,Pic
can be explained within an unshielded or weakly shielded model, all CO-rich disks 
 require substantial shielding with $\epsilon_{\rm CO}>100$, i.e. following the proposal 
 of \citet{kral2018} the latter objects can be rightfully called {\sl shielded debris disks}.

CO column densities needed for efficient self-shielding of these systems are also shown in
Fig.~\ref{fig:shielding}. By comparing these values with the vertical CO column densities 
of the given disks we found that self-shielding can play an important role only in the HD\,32297 and 
HD\,121617 systems. Even in these two disks, however, more efficient shielding is needed for 
explaining the observed CO quantity. Note that, all of our previous calculations on
shielding dealt only with $^{12}$CO molecules whose total mass was estimated from optically 
thin C$^{18}$O or $^{13}$CO observations assuming abundance ratios to be equal to the
[$^{16}$O]/[$^{18}$O] and [$^{12}$C]/[$^{13}$C] isotope ratios measured in the 
local interstellar matter \citep[see Sect.~\ref{coemission} and][]{moor2017}. 
UV absorption lines of $^{13}$CO and C$^{18}$O are only partly overlapping 
with those of $^{12}$CO making these lower abundance isotopologues more vulnerable 
to photodissociation, e.g. based on \citet{visser2009} in a disk where 
only self-shielding is taken into account the photodissociation rate of C$^{18}$O molecules -- 
depending on the $^{12}$CO column densities -- could be up to $\sim$15$\times$ higher than that 
of $^{12}$CO. By leading to higher $^{12}$CO/$^{13}$CO and $^{12}$CO/C$^{18}$O ratios this 
isotope-selective photodissociation would result in higher $^{12}$CO masses than we assumed 
in our analysis thereby further increasing the need for more efficient shielding mechanisms. 
This also means that to avoid the uncertainties associated with the abundance ratios and
 construct a more self-consistent model we rather have to reproduce 
the results of the optically thin $^{13}$CO or C$^{18}$O line observations.

We note that the currently known sample of CO-bearing debris disks may contain 
three additional young shielded debris disks. HD\,138813 and HD\,156623 exhibit 
higher $^{12}$CO line luminosities than 49\,Cet and HD\,21997.
Though, because of the lack of $^{13}$CO and/or C$^{18}$O observations, their CO mass is less constrained, 
the lower mass limits derived from the $^{12}$CO measurements imply shielding factors of 
$>$10 and $>$100 for HD\,156623 and HD\,138813, respectively. 
These results suggest that these systems likely harbor CO-rich shielded debris disks too.  
A third system, HD\,121191, harbors a CO mass of 2.5$\times$10$^{-3}$\,M$_\oplus$ \citep[][taking into 
account the new Gaia DR2 distance]{moor2017}. 
This remains below the average CO mass of the abovementioned six systems but  
almost two orders of magnitude higher than that of $\beta$\,Pic. Though the paucity of information 
on the spatial distribution of gas and dust and their relationship in this system makes the modelling 
less reliable, this disk might also require strong shielding \citep{moor2017,kral2018}. 
Similarly to the previously discussed CO-rich systems all these three disks surround
A-type star with ages $<$50\,Myr. 
  
\subsection{Origin of gas in 49\,Cet and HD\,32297}  \label{sec:gasorigin}
In the literature two models are proposed to explain the existence of shielded debris disks. 
The hybrid disk model \citep{kospal2013,pericaud2017} is motivated by the fact that 
all known representatives of this class are very young. According to this scenario, the gas 
in these systems is mostly primordial and dominated by hydrogen molecules that 
provide effective shielding for CO. The observed CO molecules could partly be primordial 
and collisions of icy bodies and grains -- that are thought to be responsible for the continuous 
replenishment of the dust component -- may also produce gas. 
Recently \citet{kral2018} proposed an alternative model in which both the dust and gas components 
are second generation. According to this scenario second generation molecules, including CO, are produced
from volatile-rich solid bodies located in a planetesimal belt. Should the CO production rate be 
sufficiently high and the viscous evolution of the gas slow enough, neutral atomic carbon gas, the 
photodissociation product of CO, {can accumulate so considerably that it becomes 
optically thick to UV radiation. Since the photoionization of C$^0$ occurs at the same 
UV wavelengths as the photodissociation of the CO molecules, the carbon component 
could shield CO efficiently.} 
This leads to the formation of a CO-rich shielded debris disk.
 \citet{kral2018} found that the disks around HD\,21997, HD\,121191, HD\,121617, HD\,131488, and HD\,131835 
can be explained with this model. 

\subsubsection{Upgraded shielded secondary gas disk model}
In the following, we evaluate whether {this} shielded secondary gas disk model can reproduce the 
observed quantities of 49\,Cet and HD\,32297 as well. 
The model presented in \citet{kral2018} can be used to model gas released from exocometary bodies and 
it is now able to follow the evolution of massive CO disks thanks to the inclusion of a new ingredient 
in their modelling: 
{C$^0$} shielding of CO. It is also the first model to follow the evolution of 
CO together with C$^0$. 
CO is originated from the cometary bodies at the parent belt location at a rate 
$\dot{M}_{\rm CO}$. CO is then initially photodissociated in $\sim100$yr when no 
shielding occurs, which forms a gas disk made of atomic carbon and oxygen. In this 
new model, \citet{kral2018} follow the evolution of carbon taking into account that 
at each time step there is some new carbon coming from CO destruction and that 
carbon depletes at a given radial location $R$ because it viscously spreads. The 
viscosity $\nu$ is parameterized with an $\alpha$ parameter such that 
$\nu=\alpha c_s H$, where $c_s$ is the sound speed 
and $H$ the scaleheight of the disk. In turn, when and if enough carbon accumulates, 
the model includes a back reaction on the CO photodissociation timescale that becomes 
longer owing to carbon shielding that can be very efficient. When the CO photodissociation 
timescale becomes longer than the viscous timescale (roughly equal to $R^2/\nu$), the 
CO gas disk has also time to viscously spread, which is included in the model.

{We note that in debris disks $\alpha$ values could be very different from those in 
protoplanetary disks for several reasons. \citet{kral2016b} found that the 
magnetorotational instability could work in a debris disk environment and 
could even lead to high $\alpha$ values because of the high ionisation 
fraction that can be reached in unshielded disks \citep[mostly made of 
ionised carbon, such as in $\beta$ Pic,][]{cataldi2018}. 
In HD\,32297 and 49\,Cet, the ionisation fraction could be much lower 
because of the strong shielding leading to drastically} {lower} {$\alpha$ 
values.} 

{Stellar radiation pressure can have a substantial effect on gas particles, 
leading to the blowout of certain species \citep{fernandez2006}. Especially 
around the more luminous 49\,Cet, carbon atoms can also be subject of 
radiation forces. However, as \citet{kral2017} demonstrated, when a disk 
 becomes optically thick to UV radiation in the radial direction, 
 the efficiency of the star's radiation pressure decreases dramatically. 
They found that in disks with a CO production rate of 
$>$10$^{-4}$\,M$_\oplus$/Myr -- which is significantly lower than in our 
targets (see Sect.~\ref{modellingresults}) -- carbon can be protected from 
expulsion. Therefore this effect was not considered 
in our modelling. }

The model we use in this paper is based on the \citet{kral2018} model described 
above but it has been improved. Now, instead of just following the evolution at 
the parent belt location, we follow the evolution at all radial locations. This 
new 1-D model (which will be presented in detail in a forthcoming paper) thus 
solves in parallel the viscous diffusion equations \citep[see][and also how we 
proceeded in \cite{kral2016}]{lynden-bell1974} for CO and C$^0$ together with 
the abovementioned source and sink terms. We also follow the viscous evolution 
of $^{13}$CO and C$^{18}$O (which we assume to be released at the ISM 
proportion compared to $^{12}$CO) together with the fact that self-shielding 
differs for these different isotopes \citep{visser2009}. We therefore take 
into account the isotope selective photodissociation for the first time in a 
debris disk model. Another upgrade is that this new 1-D code calculates the 
ionisation fraction in the disk at every radial location taking into account 
the UV radiation from the interstellar medium as well as the star. This code 
is thus ideally suited to model disks where spatial information on the gas and 
dust locations have previously been obtained from resolved ALMA observations. 
We will now describe how we used it for the specific systems around 49\,Cet 
and HD\,32297.

\begin{figure*}
\begin{center}
\includegraphics[angle=0,scale=0.39]{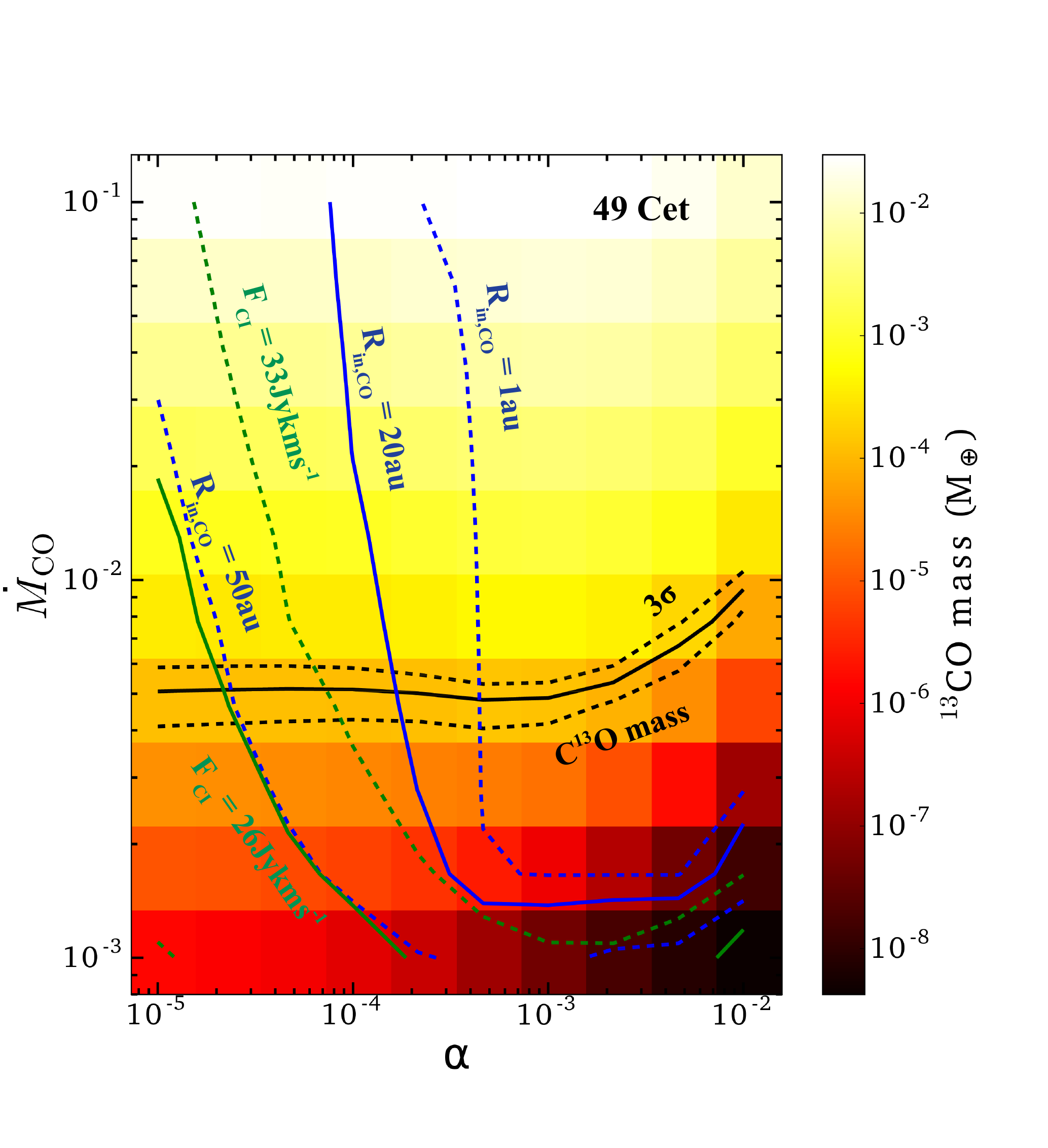}
\includegraphics[angle=0,scale=0.39]{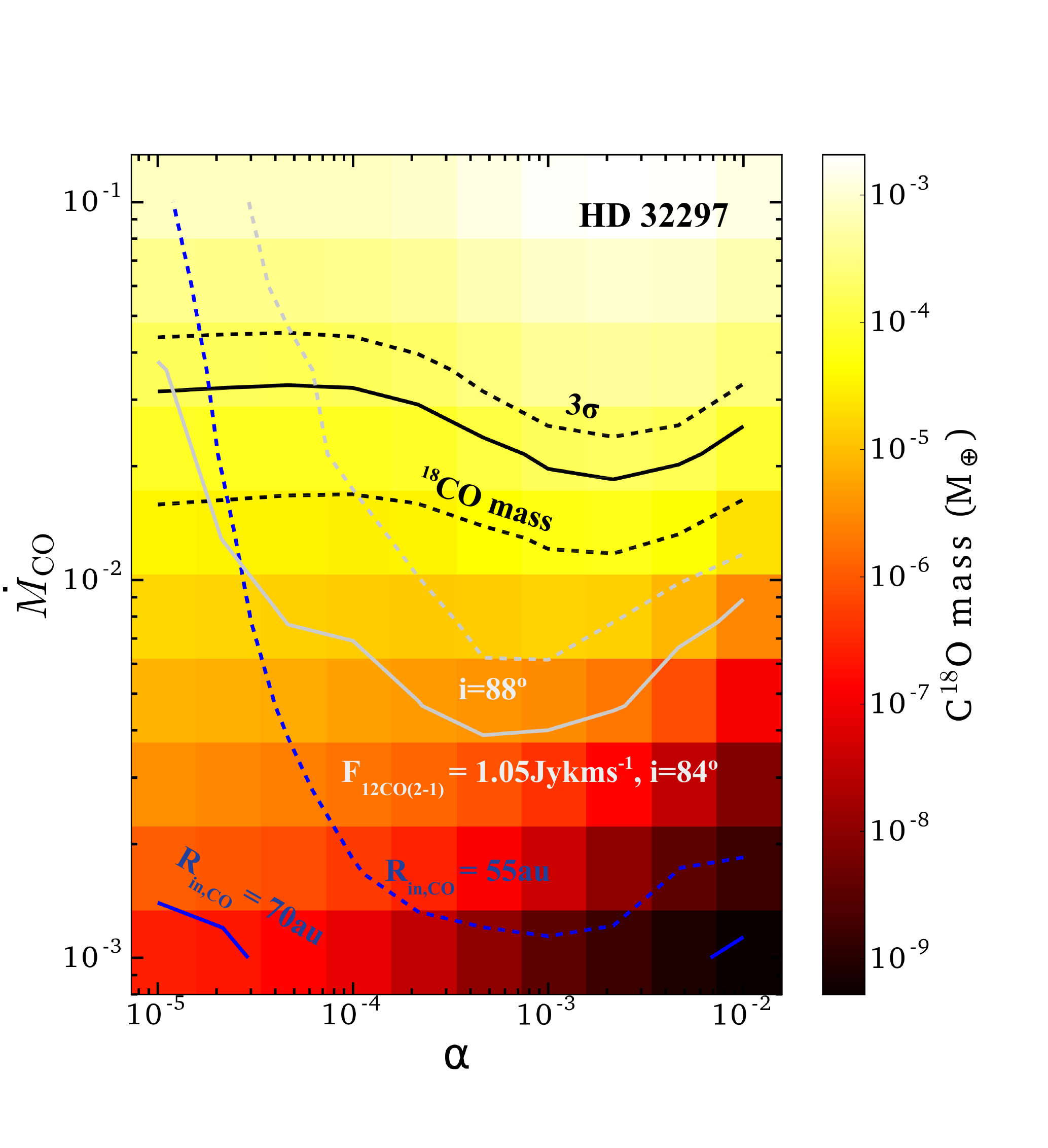}
\end{center}
\caption{ {$^{13}$CO (49\,Cet, left) and C$^{18}$O (HD\,32297, right) gas masses derived 
in our secondary gas disk models for different CO production rates $\dot{M}_{\rm CO}$ 
($M_\oplus {\rm Myr^{-1}}$) and different $\alpha$ viscosities.
Solid black contours correspond to the observed $^{13}$CO and C$^{18}$O masses, while 
the black dashed contours mark the 3$\sigma$ uncertainties. Blue lines denote models with different 
inner CO disk radii ($R_{\rm in,CO}$). 
Solid and dashed green lines in the left panel outline models 
producing the measured CI line flux and its higher 3$\sigma$ uncertainty.
The light gray lines in the right panel show models reproducing the observed $^{12}$CO line flux 
assuming disk inclinations of 84{\degr} (solid line) and 88{\degr} (dashed). In the case of 49\,Cet 
the acceptable region in the parameter space is defined by the 3$\sigma$ upper limit of the CI line flux 
(green dashed line at 33\,Jykms$^{-1}$) and by the 3$\sigma$ upper limit of $R_{\rm in,CO}$ 
(blue dashed line at 50\,au) and 
by the $\pm$3$\sigma$ uncertainties of the $^{13}$CO mass (dashed black lines). 
In the case of HD\,32297 the acceptable region is defined by the 3$\sigma$ lower limit of the 
$R_{\rm in,CO}$ (blue dashed line at 55\,au) and by the $\pm$3$\sigma$ uncertainties of the 
C$^{18}$O mass (dashed black lines).} 
\label{fig:modelling}}
\end{figure*}

\subsubsection{Modelling results} \label{modellingresults}
In the course of modelling we assumed that the collisional cascade and thereby the production of 
secondary CO gas is initiated 5\,Myr after the birth of the systems and then our simulations are 
run for 40\,Myr and 25\,Myr for 49\,Cet and HD\,32297, respectively. 
For HD\,32297, we adopted a planetesimal belt between 80 and 120\,au, while in the case of 
49\,Cet we used a disk model with an inner and outer radii of 71\,au and 153\,au, respectively (see Sect.~\ref{sec:shielding}). 
We took a uniform gas temperature of 20\,K {(Sect.~\ref{coemission})} in the whole disks.
In both cases we hypothesized that collisions between planetesimals 
result in fragmentation in the whole disk throughout our simulation, i.e. the disk 
is pre-stirred \citep{wyatt2008}. When releasing the gas, we initially assume a constant mass input rate with 
radial distance not to favor any gas release mechanism over others.
To find a best fit model to the $^{13}$CO and C$^{18}$O masses derived from observations for 49\,Cet and HD\,32297, 
we run a large grid of 100 models to explore the parameter space in $\dot{M}_{\rm CO}$ and $\alpha$. The grid is 
logarithmic for both parameters and comprised of 10 elements going from $10^{-3}$ to $10^{-1}$\,M$_\oplus$/Myr for 
$\dot{M}_{\rm CO}$ and from $10^{-5}$ to $10^{-2}$ for $\alpha$. For each model, we compute the total $^{13}$CO 
(for 49\,Cet) or 
C$^{18}$O (for HD\,32297) masses and compare them to observed values. Masses found by the model for each part of the 
parameter space are shown in Figure~\ref{fig:modelling}. 
Solid black contours correspond to the observed $^{13}$CO and C$^{18}$O
masses, while the black dashed contours mark the 3$\sigma$ uncertainties (Table~\ref{tab:imageparams}).

For 49\,Cet (left panel), we found that models with $\dot{M}_{\rm CO}$ ranging 
from 0.004 to 0.01\,M$_\oplus$/Myr and with any $\alpha$ can explain the measured $^{13}$CO mass. 
To further constrain the acceptable parameter space we need to consider other 
observational properties. 
By combining ALMA CO (3--2) observations with previous SMA CO (2--1) data,  
  \citet{hughes2017} derived an inner disk radius of 20\,au as a best fit. 
Detecting \ion{C}{1} $^3$P$_0$--$^3$P$_1$ emission (at rest frequency of 492.161\,GHz) towards 49\,Cet
with the single dish ASTE radio telescope, \citet{higuchi2017} derived 25.8$\pm$2.3\,Jykms$^{-1}$ for the integrated 
line flux\footnote{\citet{higuchi2017} quoted the line flux in unit of 
Kkms$^{-1}$ (0.45$\pm$0.04\,Kkms$^{-1}$). To convert K to Jy we followed the description at \url{https://alma.mtk.nao.ac.jp/aste/cfp2012/note.html} adopting a half-power beam
width of 17{\arcsec} based on \citet{higuchi2017}.}. 
Models with different $\alpha$ and $\dot{M}_{\rm CO}$ result in quite different inner CO disk 
radii ($R_{\rm in,CO}$) and CI emissions. 
The solid blue line in Fig.~\ref{fig:modelling} (left) corresponds to models 
where $R_{\rm in,CO} = 20$\,au, the best fit value suggested by \citet{hughes2017}, 
while the solid green marks models which reproduce the measured 
CI integrated line flux. In the computation of CI model fluxes we used the LIME radiation transfer 
 code \citep{brinch2010} assuming LTE conditions.
The dashed blue and green lines denote models are consistent {at a 3$\sigma$ level} with 
the observed parameters. 
{We found that there exists a region in the parameter space, 
at $2\times10^{-5} \lesssim \alpha \lesssim 9\times10^{-5}$ and $0.004 \lesssim \dot{M}_{\rm CO} \lesssim 
0.006$\,M$_\oplus$/Myr, where all three 
observational constraints, 
represented by the blue (inner CO disk radius), green (CI emission), and black ($^{13}$CO mass) lines, 
are simultaneously satisfied within their 3$\sigma$ uncertainties (dashed lines).
As a representative grid point for the subsequent simulation, we adopted 
 $\dot{M}_{\rm CO} \sim 4.6\times10^{-3}$\,M$_\oplus$/Myr and} {$\alpha \sim 4.6\times10^{-5}$.}

As Fig.~\ref{fig:modelling} (right) demonstrates, in the case of HD\,32297 
the observed C$^{18}$O mass can be reproduced with CO production rates between 0.01 and 0.04\,M$_\oplus$/Myr. 
Although \citet{macgregor2018} presented no 
detailed model for the spatial distribution of $^{12}$CO, they argued that the gas and dust components 
in this disk are co-located. For the inner edge of the dust disk they inferred a radius of 
78.5$\pm$8.1\,au. 
Fig.~\ref{fig:modelling} (right) shows models with 70\,au 
and 55\,au inner CO disk radii represented by solid and dashed blue lines, 
respectively. 
The latter value can be considered as a lower limit for the 
inner radius. These models suggest a constraint of $\alpha\lesssim$2$\times$10$^{-5}$. 
Models with $R_{\rm in,CO}=70$\,au has no intersection 
with the black line in the explored parameter space.
Furthermore we also displayed the loci of models (gray lines) that 
produce $^{12}$CO line flux corresponding to the observed value 
(1.05\,Jykms$^{-1}$, Table~\ref{tab:imageparams}). In the calculations 
we assumed two different inclination values, 88{\degr} (dashed line) and 
84{\degr} (solid), the best fit parameters inferred from scattered light 
\citep{boccaletti2012} and continuum millimeter \citep{macgregor2018} 
observations of the disk. 
Model line flux computations were performed with the LIME tool using LTE approach.
Intersection of the latter loci with the 
black lines is broadly consistent with the $R_{\rm in,CO}=55$\,au models. 
The grid point located closest to these crossings 
has $\dot{M}_{\rm CO} \sim 3.6\times10^{-2}$\,M$_\oplus$/Myr and 
$\alpha \sim 1.5\times10^{-5}$ for HD\,32297.

\begin{figure*}
\begin{center}
\includegraphics[angle=0,scale=0.28]{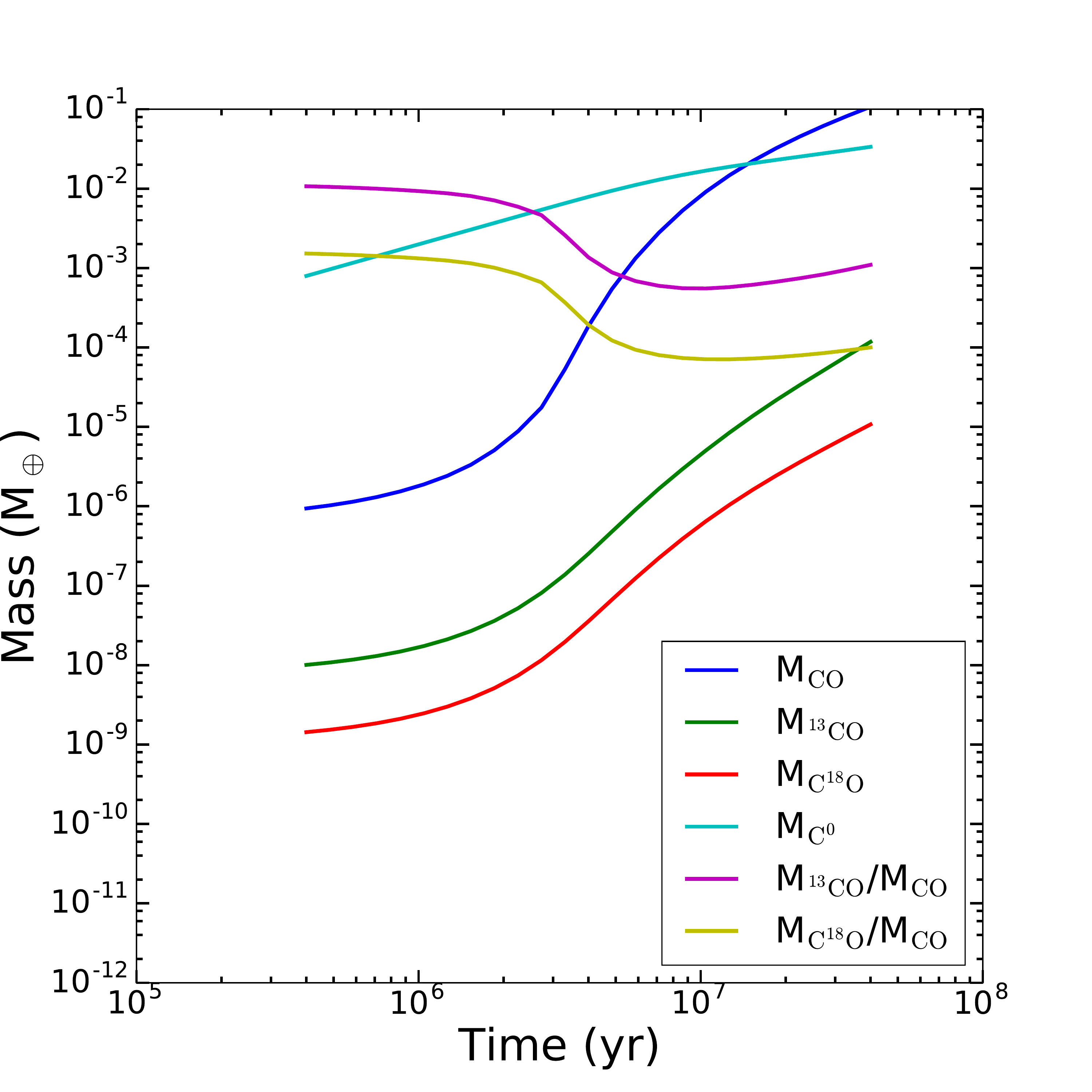}
\includegraphics[angle=0,scale=0.28]{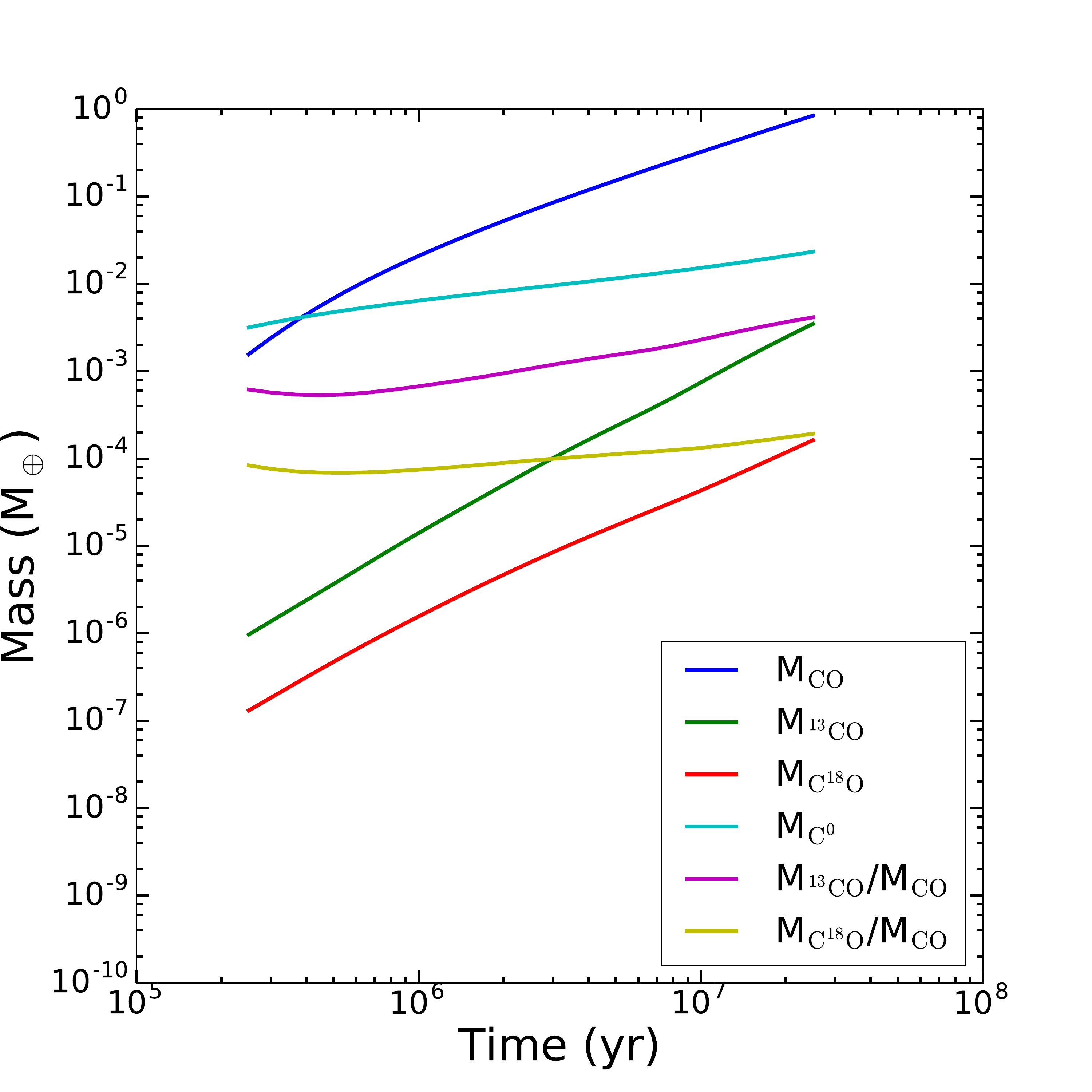}
\end{center}
\caption{ {The mass evolution of $^{12}$CO, $^{13}$CO, C$^{18}$O gas and C$^0$ gas components in
representative secondary gas disk models of 49\,Cet (left) and HD\,32297 (right).
Variation of the $M_{^{13}{\rm CO}}$/$M_{^{12}{\rm CO}}$ and $M_{\rm C^{18}O}$/$M_{^{12}{\rm CO}}$ 
mass ratios during the disk evolution are also shown. The models predict substantial variations in 
the mass ratios. There is an initial (very short for HD~32297) period when the mass ratios decrease. 
The $^{12}$CO line becomes gradually optically thick leading to stronger self-shielding of the
$^{12}$CO molecules. At the same time the optically thin rarer isotopologues remain 
less efficiently shielded (isotope selective photodissociation). This leads 
to their decreasing mass ratios with respect to $^{12}$CO.
Later, the shielding of $^{13}$CO and C$^{18}$O also becomes more efficient due to the increasing 
carbon and CO column densities and the mass ratios start to increase.}  
\label{fig:modelling2}}
\end{figure*}

We note that in the case of HD\,32297 the agreement between the model and 
data is more marginal than for 49 Cet and requires very low $\alpha$ values.  
{As we mentioned earlier, because of strong shielding, the ionization 
fraction of this disk could be significantly lower than in the more 
tenuous gas disk of $\beta$~Pic.}
The lower ionisation fraction will also favour ambipolar diffusion 
that could become important in these low gas density disks. Using eq.~4 in 
\citet{kral2016b}, we calculate that the ambipolar parameter for carbon is 
$\sim 100 (\frac{T}{20{\rm K}})^{-1/2} (\frac{\Sigma_n}{10^{-5}{\rm g/cm}^2}) (\frac{f}{10^{-2}})$, 
where $T$ is the gas temperature, $\Sigma_n$ the neutral carbon density and 
$f$ the ionisation fraction of carbon. We see that indeed, in HD\,32297, for 
ionisation fraction lower than about 10$^{-2}$ \citep[which is typical for 
shielded disks as shown in][]{kral2018}, the ambipolar parameter becomes lower 
than 100 and thus the ions and neutrals are not well coupled, which provides 
very low $\alpha$ values in MRI-simulations 
\citep[see][and references therein]{kral2016b}.

Our best fitting secondary gas models predict CO production rates of $\sim$0.005\,M$_\oplus$/Myr and 
$\sim$0.035\,M$_\oplus$/Myr for 49\,Cet and HD\,32297, respectively. Considering the 
estimated mass loss rates, 0.1\,M$_\oplus$/Myr for 49\,Cet and 5.2\,M$_\oplus$/Myr for HD\,32297 
(see Sect.~\ref{sec:shielding}), these $\dot{M}_{\rm CO}$ rates can be reproduced with CO+CO$_2$ ice mass 
fractions of {$\gamma \sim 5$\% and $\sim$0.7\%, respectively}.
Using solar system comets as analogues, where 
this fraction range between 2\% and 27\% \citep{mumma2011,matra2017a}, the required $\gamma$ 
values are not unrealistically high. 
Considering the obtained CO mass production rates and the adopted evolutionary 
times, the total CO gas mass release of the disks around 49\,Cet and HD\,32297 
 throughout their evolution is $\sim$0.2\,M$_\oplus$ and $\sim$0.9\,M$_\oplus$, respectively.

\begin{figure*}
\begin{center}
\includegraphics[angle=0,scale=0.28]{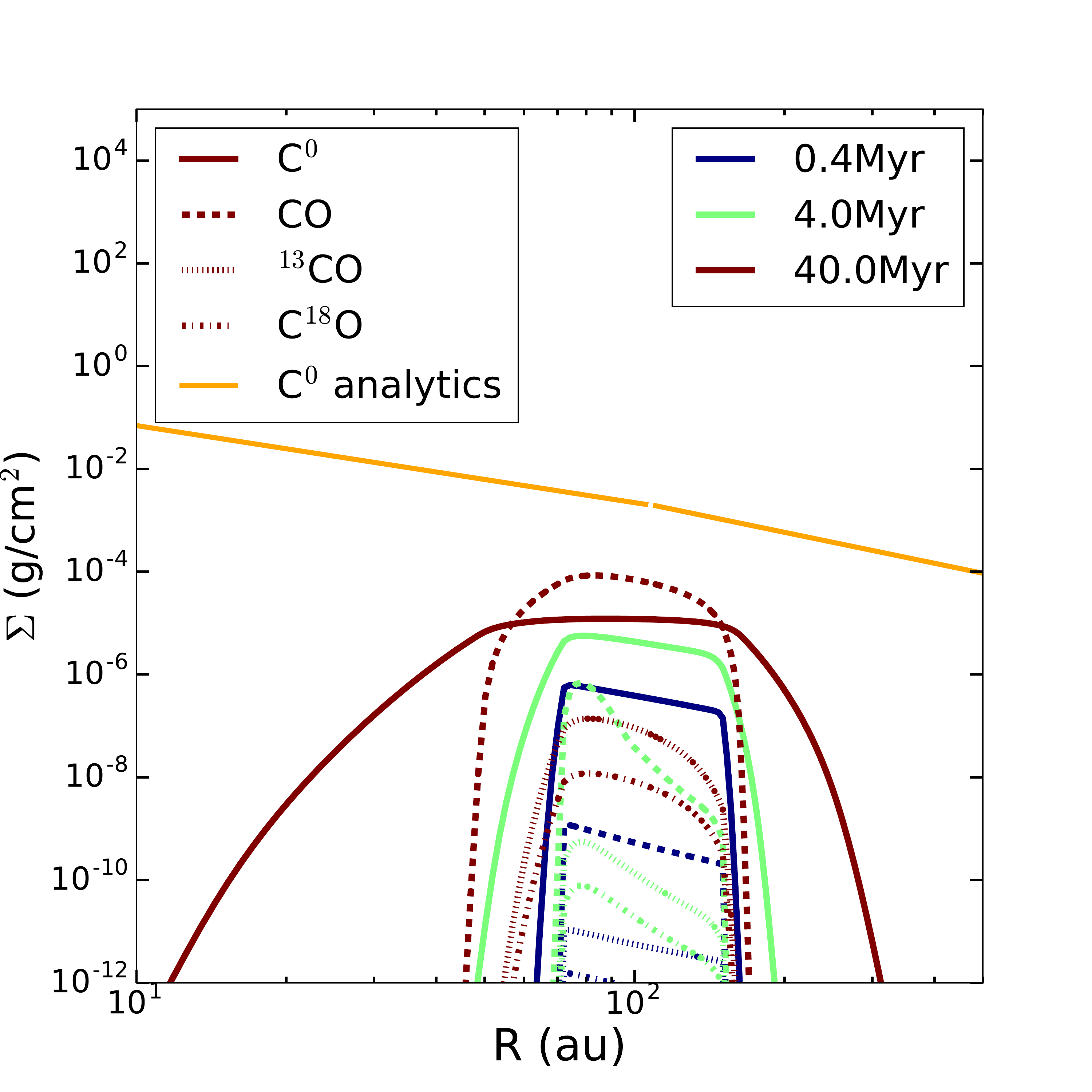}
\includegraphics[angle=0,scale=0.28]{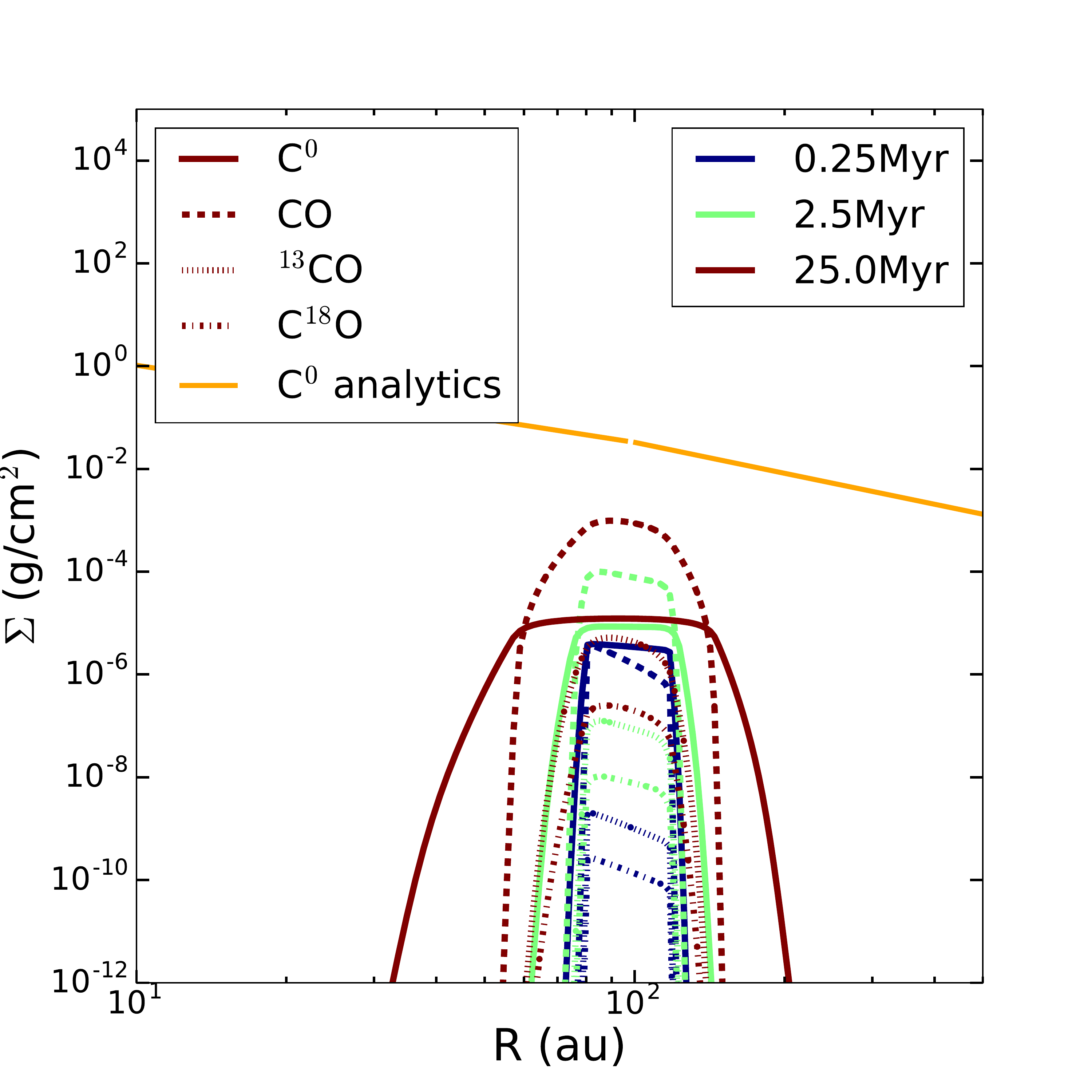}
\end{center}
\caption{
{Radial surface density distributions of $^{12}$CO, $^{13}$CO, C$^{18}$O molecules and 
C$^0$ atoms at three different evolutionary phases in
the best fit secondary gas disk models of 49\,Cet (left) and HD\,32297 (right). 
The C$^0$ analytics line (orange) is the level C$^0$ would reach at steady state for 
the given $\dot{M}_{\rm CO}$ and $\alpha$ values without shielding (assuming an 
ionisation fraction of zero), see eq.~1 in \citet{kral2018}.}
\label{fig:modelling3}}
\end{figure*}

\subsubsection{Mass evolution of the gas components}
Figure~\ref{fig:modelling2} displays the mass evolution of the $^{12}$CO, $^{13}$CO, 
C$^{18}$O gas, and carbon components in the best fit secondary gas disk models of 
49\,Cet and HD\,32297. According to this, the disk of HD\,32297 reaches a CO mass of $\sim$0.01\,M$_\oplus$ after 
$\sim$0.5 million years of evolution and then appears as a CO-rich debris system for nearly 
all of its lifetime. In the 49\,Cet system it takes $\sim$12\,Myr to reach the same level of CO content.   
As these {plots} illustrate, the $M_{^{13}{\rm CO}}$/$M_{^{12}{\rm CO}}$ and $M_{\rm C^{18}O}$/$M_{^{12}{\rm CO}}$ 
mass ratios show 
a significant variation during the studied evolutionary period. These changes are related to 
the isotope selective photodissociation. As the $^{12}$CO line becomes more and more optically thick and 
the column density of neutral carbon increases, the $^{12}$CO line acquires strong shielding 
while the photodissociation timescale of rarer isotopologues, that remain optically thin, remains short. 
This leads to decreasing mass ratios.
Then, as the column densities of neutral carbon, 
$^{13}$CO, and C$^{18}$O increases further, and thus the shielding of less abundant CO isotopologues 
become more efficient, the mass ratios start to increase but do not reach the typical values found for the 
local interstellar matter (that we initially assumed for our icy planetesimals) even at the end of the studied period. 
As a result, the $^{12}$CO masses we obtain 
in these simulations (0.1\,M$_\oplus$ for 49\,Cet and 0.9\,M$_\oplus$ for HD\,32297) are 
{about an order of magnitude} higher than we did in Sect.~\ref{coemission}.
In the case of HD\,32297 the evolution is so rapid that the initial decreasing trend is less clearly outlined and 
significantly shorter than for 49\,Cet.
While in HD\,32297 the CO mass exceeds that in 49\,Cet by a large factor, our model predicts very similar C$^0$ 
masses of 0.03 and 0.04\,M$_\oplus$ for HD\,32297 and 49\,Cet, respectively. These values are similar to the 
C$^0$ mass found around HD\,131835 \citep{kral2018}.

\subsubsection{Evolution of gas surface density distribution}
Figure~\ref{fig:modelling3} shows the actual radial surface density distributions of the $^{12}$CO, 
$^{13}$CO, and C$^{18}$O molecules as well as of carbon atoms for {three} different evolutionary 
times using the best fit model of the specific target. 
We predict that in both systems the radial distribution of C$^0$ is more extended than those of 
the different CO molecules. {This is because when the carbon gas spreads viscously away from the 
gas production site, the most displaced regions will become too thin 
to shield CO molecules. Drastic depletion of CO through photodissociation 
in these zones then results in a larger C$^0$ to CO width.}
Future high angular resolution mapping of CI emission will make it   
possible to check these predictions and give further constraints on the models \citep[see][]{kral2018}.  

\subsubsection{A possible primordial origin of the gas component}
We found that the shielded secondary gas disk model offers a viable explanation 
for the prominent CO content of the young debris disks around 49\,Cet and HD\,32297.
Although this finding implies that there is no need for additional 
large amount of primordial H$_2$ molecules to explain the shielding of CO gas, 
we cannot rule out their presence based on 
our current results. To clarify this question, further observations are needed.
One possible way to constrain the overall composition of the gas is to derive 
the scale height of the disk that depends on the gas temperature 
and the mean molecular weight \citep{hughes2017,kral2018}. By measuring the
kinetic gas temperature and the vertical extent of the disk thus we can 
estimate the mean molecular weight that is expected to be very different 
in the primordial/secondary gas scenarios due to the different gas 
composition. \citet{hughes2017} has already applied this 
method for 49\,Cet. Though the vertical size of the disk remained unresolved 
in their observation, the obtained upper limit favors a higher mean
molecular weight, i.e. a secondary gas composition. Due to its nearly 
edge-on orientation and brightness in CO lines the disk around HD\,32297 
is ideal for such observation. Though the detection of cold H$_2$ gas 
is very difficult in emission, its electron transitions can be observed 
in absorption at far-ultraviolet wavelengths offering an opportunity 
to measure the H$_2$ column density in disks seen close to edge-on. 
By analyzing spectroscopic observations obtained with the FUSE ({\sl Far Ultraviolet Spectrocopic Explorer}) 
satellite, 
\citet{lecavalier2001} were able to set an upper limit of $<$10$^{18}$\,cm$^{-2}$ on the column density 
of H$_2$ molecules in the gaseous edge-on disk of $\beta$\,Pic. Comparing this result to 
the column density of CO molecules detected in absorption resulted in a CO to H$_2$ ratio of
$>$6$\times$10$^{-4}$ indicating -- in good accordance with conclusions from other observations 
\citep{fernandez2006,kral2016,matra2017a,cataldi2018} -- 
that the gas is of second generation in $\beta$\,Pic. Later, \citet{martinzaidi2008} reported an even 
stricter upper 
limit of $N_{\rm H_2}<2.6\times$10$^{17}$\,cm$^{-2}$
for $\beta$\,Pic. No similar data are available for any of the shielded debris disks.
Among them, HD\,32297 harbors the most edge-on disk 
($83.6< i <88{\degr}$)
offering a special opportunity 
to directly measure the CO/H$_2$ ratio using the same method in the future.
For outlines of additional possible observational methods that could be used to further constrain 
the origin of gas in shielded debris disks we direct the reader to \citet{kral2018}.

\subsection{Gas-dust interaction}

Gas, if presents in sufficient amount, can have a significant influence on the 
dynamics of small debris particles \citep[e.g.][]{takeuchi2001,besla2007,
richert2018}. With their protoplanetary level of CO gas, the debris disks 
of 49\,Cet and HD\,32297 are among the best candidates where gas-dust 
interactions can really manifest. \citet{kral2018} pointed out that such
interactions could be important in the CO-rich disk around HD\,131835 \citep[see also][]{feldt2017}. 
In order to assess how well the gas 
and grains are coupled in 49\,Cet and HD\,32297, we estimated the dimensionless 
stopping time that shows the timescale (in orbital units) needed for a dust 
particle to be entrained by the gas. According to \citet{richert2018} the 
dimensionless stopping time (a.k.a. the Stokes number, $St$) can be computed 
as:
\begin{equation}
St \approx \frac{1}{\beta} \bigg( \frac{\Sigma_{\rm g}}{9\times10^{-5} 
{\rm g cm^{-2}}}\bigg)^{-1} \bigg(\frac{L_*}{L_\odot}\bigg) 
\bigg( \frac{M_*}{M_\odot} \bigg)^{-1}
\label{eq:stokes}
\end{equation} 
where $\beta$ is the ratio of radiation to gravitational force, 
$\Sigma_{\rm g}$ is the gas surface density, while $L_*$ and ${M_*}$ are the 
luminosity and mass of the central star. 
Smaller Stokes number means stronger coupling between dust and gas.
Assuming parent bodies on a circular orbit, $\beta$ is equal to 0.5
for the smallest bound grains \citep{krivov2010}. 
Concerning $\Sigma_{\rm g}$, we took the highest total (C$^0$+CO) gas 
surface density of our secondary gas disk models (Fig.~\ref{fig:modelling3}). 
Then Eq.~\ref{eq:stokes} yields $St\sim$20 and $\sim$1
for the smallest bound grains in 49\,Cet and HD\,32297, respectively.
These values, however, should be considered as upper limits
since additional daughter products, H, O, from photodissociation of CO, CO$_2$, and 
H$_2$O have not been taken into account in 
this calculation. In solar system comets the H$_2$O/CO mass ratio ranges 
between 3 and 200 \citep{mumma2011}. Considering these additional products, like H and O, 
the $St$ numbers could be at least a few times lower. This certainly results in 
$St \lesssim 1$ for HD\,32297, indicating that the smallest bound grains are at 
least marginally coupled to gas via drag even if the icy parent bodies have 
very low H$_2$O/CO ratio. In the 49\,Cet system more H$_2$O daughter products, thus 
more water-rich planetesimals are needed to reach the same level of gas-dust coupling for 
such grains. Assuming compact spherical grains with radiation pressure efficiency 
of unity, the $\beta$ parameter can be expressed as $\beta = 0.574 \big(\frac{L_*}
{L_\odot}\big) \big( \frac{M_*}{M_\odot} \big)^{-1} \big( \frac{1\mu m}{s} \big) \big(\frac{1 
{\rm g cm^{-3}}}{\rho} \big)$, where $s$ and $\rho$ are the size and density 
of the grains \citep{burns1979}. Taking a grain density of 2\,g~cm$^{-3}$, 
the grain size of $\beta$=0.5 particles (blowout grain size) in the 49\,Cet and 
HD\,32297 systems are $\sim$5{\micron} and $\sim$3{\micron}.

The presence of gas has an influence on the dynamics 
of unbound grains with $\beta>0.5$ as well. While they would normally escape 
from the system on the local dynamical timescale, their interaction with the 
gas, e.g. gas drag, can result in longer lifetimes leading to an accumulation of very small 
particles compared to a gas-free situation \citep{lieman-sifry2016,kral2018,richert2018,wyatt2018}. 
In both of our targets 
there are indications that trapping of small grains is really happening.
Analyzing 12.5{\micron} and 17.9{\micron} Keck images of 49\,Cet, \citet{wahhaj2007} 
argued that the bulk of the observed mid-IR emission comes from grains with 
size of $\sim$0.1{\micron} located between 30 and 60 AU from the star. This overlaps with the region where 
gas was also found \citep{hughes2017}. Interactions with this gas material can 
explain how such small particles, well below the blow-out limit, can be retained     
in the disk. 

Based on SED analysis \citet{donaldson2013} derived a characteristic 
dust temperature of 83\,K for the cold outer component of HD\,32297. Supposing that 
dust grains act like blackbodies, this temperature would correspond to a dust ring 
located at a radius of 32\,au from the star 
\citep[$R_{\rm bb} ({\rm au}) = \big(\frac{L_*}{L_{\sun}} \big)^{0.5} \big(\frac{278.3\,{\rm K}}{T_{\rm dust}} \big)^2$,][]{backman1993}. 
High angular resolution millimeter image of the source, however, implies a substantially 
larger radius for the outer belt that was modelled by 
a planetesimal belt encompassed by a very extended halo \citep{macgregor2018}. The radius of this 
planetesimal belt component 
is $\sim$100\,au (average of the 
best-fit inner and outer belt radii). From this radius we obtain a $\Gamma = R_{\rm disk}/R_{\rm bb} 
= T_{\rm disk}^2/T_{\rm bb}^2 $ ratio of 3.1 that, 
because of the presence of the halo, 
should be considered as a lower limit. \citet{pawellek2014} and \citet{morales2016} measured lower $\Gamma$ ratios 
 for other resolved debris disks around stars with similar luminosity (6--10\,L$_\sun$), indicating 
 that the outer disk of HD\,32297 is unusually warm. The larger average dust temperature might be attributed 
 to the presence of warm micron-sized particles that may have longer lifetime than in gas-poor debris 
 disks around similarly luminous host stars.

 High angular resolution mid-IR {thermal emission and near-IR scattered light images} of HD\,32297 provide 
 further evidence for the presence of small dust. Based on their 11.2{\micron} observation, \citet{fitzgerald2007} 
 argued that warm dust grains emitting at this wavelength are situated beyond a radial separation of 
 $\gtrsim$0.65{\arcsec} 
 ($\gtrsim$86\,au), thus are co-located with the outer gas and dust belt mapped in the millimeter 
 regime \citep{macgregor2018}. 
 By analyzing their 11.7{\micron} and 18.3{\micron} images of the source, \citet{moerchen2007} 
 also found that the 
 emitting grains are mostly concentrated in the outer regions ($\gtrsim$94\,au, considering the new Gaia DR2 distance) and 
 derived typical dust temperature of $\sim$180-190\,K. Both works concluded that small 
 submicron sized particles were present at large distances from the star. Noteworthy this also means that the warm component
 inferred from the SED analysis \citep{donaldson2013} cannot be exclusively linked to an inner 
 dust belt. 
 {By analyzing their multiband scattered light data on HD\,32297 
\citet{bhowmik2019} measured a blue color in the near-infrared spectrum
that was interpreted as a signature of a significant presence of 
 grains smaller than the blowout limit.}

Several other CO-bearing debris disks show signatures of considerable amount of small dust particles. By modelling 
resolved debris disks in their survey \citet{lieman-sifry2016} found evidence for the presence of grains smaller 
than the blowout size in two gaseous debris disks, HD\,138813 and HD\,156623. \citet{moor2017} discovered that the disks of 
HD\,121617 and HD\,131488 -- similarly to HD\,32297 -- exhibit outstandingly high $\Gamma$ factors, 
indicative of warm, small solids. The spatially resolved mid-IR image of HD\,131835 also revealed very hot, 
supposedly submicron sized particles via their strong emission in the outer disk \citep{hung2015}. Thus, 
the existence of small grains in CO-rich debris disks seems to be a common phenomenon.

Due to gas drag, grains can move radially inward or outward in the disk. The direction, speed, and degree of this radial 
migration depend on many different factors, such as the density and temperature distribution of the gas component, 
the size of the particle, the strength of the radiative forces, and the collisional timescales. 
Grains at the lower end of the size distribution typically migrate outward  \citep{takeuchi2001} 
potentially leading to a detachment between the ring of parent planetesimals and the small grain population. 
In a recent model of gas-dust interactions in optically thin disks, \citet{richert2018} considered dust-gas drag with backreaction, 
photoelectric heating of gas by dust, and stellar radiation pressure of dust. In their simulations, they detected 
various structures like concentric rings, arcs, spirals arms in the dust distribution, attributed to photoelectric instability.
Though their results cannot be directly applied to our cases (e.g. they adopted a G-type star as central object resulting 
in substantially weaker radiation field), in their models with $St \lesssim 1$ they saw several of the abovementioned features.
We assume that in the case of HD\,32297, where the Stokes number for $\beta=0.5$ dust particles is also low ($St \lesssim 1$), 
the same physical processes like gas drag driven detachment and photoelectric instability could be operational. 
The resulting structures, if any, can be best observed in the distribution of small grains outlined by high spatial 
resolution scattered light images as pointed out in \citet{kral2018}.

So far we supposed that our two targets harbor secondary gas disks. As we already noted in 
Sect.~\ref{sec:gasorigin}, it cannot be excluded that the detected gas component is the residual of 
the primordial disk whose mass is dominated by H$_2$ molecules. 
Adopting the canonical H$_2$/CO abundance ratio of 10$^4$ (i.e. a mass ratio of $\sim$700), 
the gas surface densities in our two disks could be several hundred times higher, meaning 
that even the $>$100{\micron} sized dust particles are well coupled to the gas.
From the analysis of the 850{\micron} continuum ALMA image of 49\,Cet, \citet{hughes2017} found 
a single power-law disk model with a ring-like enhancement at 110\,au in their best fit model.
If the gas component is primordial, this dust ring could be attributed to gas-dust interactions.


\section{summary} \label{sec:summary}

We used the ALMA 7m-array to measure the $^{13}$CO and C$^{18}$O isotopologues towards young debris disks 
around 49\,Cet and HD\,32297. Our observations revealed unexpectedly high CO gas contents comparable to 
the levels in protoplanetary disks. 
Though collisions and evaporation of icy bodies in a debris disk 
can naturally lead to gas production, the presence of the observed huge amount of CO gas in these systems 
needs a special environment in which the released CO molecules are shielded against 
UV photons from the stellar and interstellar radiation field. 

Previous secondary gas models 
were not able to account for debris systems with such high CO masses putting forward the concept 
of hybrid disks where the gas component is dominated by residual primordial H$_2$ molecules.
To examine whether the disks of 49\,Cet and HD\,32297 could be fully secondary we utilized 
the shielded secondary gas model proposed recently by \citet{kral2018}. 
We found that by assuming a CO production rate of $\sim$0.005-0.03\,M$_\oplus$/Myr and slow viscous 
evolution with $\alpha$ of $< 10^{-4}$, shielding by carbon atoms and self-shielding by CO 
together could be so effective that the obtained $^{13}$CO (for 49\,Cet) and C$^{18}$O (for HD\,32297) 
masses can be explained in the framework of this model. 
In the case of 49\,Cet the line flux of the predicted \ion{C}{1} 
component was also found to be consistent with the observed one. 
The necessary CO production rates in the two systems are consistent with their dust production rates 
derived from observations, if we adopt ice mass fractions typical in solar system comets. 
Moreover, the total CO mass released from the planetesimals over the lifetime 
of the disks do not exceed the CO content of known nearby protoplanetary disks 
around Herbig Ae stars. 

 In the presence of sufficiently dense gas, dust particles and  
gas can interact with each other in various ways. Based on gas    
surface densities derived from our new shielded secondary gas models 
we argue that small grains with sizes close to the blowout limit 
in the disk of HD\,32297 are at least moderately coupled to the gas 
and likely subject to such interactions. In accordance with this, 
several observational results indicate that the dust material of 
the gaseous outer ring of HD\,32297 is unusually warm suggesting an 
overabundance of small solids in this region. The high gas mass of HD\,32297
makes this source one of the most promising targets to study the possible outcomes 
of gas-dust interactions in optically thin dust disks. 
Though based on our secondary gas model the gas-dust coupling might be weaker 
in the 49\,Cet system, there are also indications for the presence of considerable 
amount of particles with size smaller than the blowout limit. The retention 
of these particles suggests that gas-dust interaction can happen at some level 
in this system, too. 
We note that if the gas has a primordial origin then even large grains with size of 
$>$100{\micron} could be entrained by the gas.

49\,Cet and HD\,32297 are in many ways similar to HD\,21997, HD\,121617, HD\,131488, and 
HD\,131835. The central stars in these systems are younger than 50\,Myr, have A spectral type, 
and have no known stellar companion. Their debris disks have dust-rich ($f_{\rm d}>5\times10^{-4}$) 
outer belts at radii larger than 30\,au. 
These belts are at least partly co-located with a gas component, 
having a protoplanetary disk like CO mass of $M_{\rm CO} > 0.01$\,M$_\oplus$.  
Though the exceptional gas content of all six disks can be explained 
in the framework of the shielded secondary disk scenario, 
we cannot rule out that in some of them 
the very effective shielding of CO is due to leftover primordial H$_2$ gas. 
Future observations are needed for a final clarification of this question.

\acknowledgments

We thank the anonymous referee for providing insightful comments that
helped us to improve the quality of this paper. We also thank 
Alycia Weinberger, Meredith MacGregor, and Gerrit van der Plas for sharing 
their results with us prior to publication.
This paper makes use of the following ALMA data:
ADS/JAO.ALMA\#2017.2.00200.S. ALMA is a partnership of ESO
(representing its member states), NSF (USA) and NINS (Japan), together
with NRC (Canada) and NSC and ASIAA (Taiwan) and KASI (Republic of
Korea), in cooperation with the Republic of Chile. The Joint ALMA
Observatory is operated by ESO, AUI/NRAO and NAOJ. 
This work has made use of data from the European Space Agency (ESA) mission
{\it Gaia} (\url{https://www.cosmos.esa.int/gaia}), processed by the {\it Gaia}
Data Processing and Analysis Consortium (DPAC,
\url{https://www.cosmos.esa.int/web/gaia/dpac/consortium}). Funding for the DPAC
has been provided by national institutions, in particular the institutions
participating in the {\it Gaia} Multilateral Agreement.
Our work was supported by 
the Hungarian OTKA grants K119993 and KH130526.

\vspace{5mm}
\facilities{ALMA}

\software{CASA \citep[v4.7.2][]{mcmullin2007}, LIME \citep{brinch2010}}

\bibliographystyle{apj}



\end{document}